\documentclass[12pt]{article}

\usepackage{amssymb,amsfonts,amsmath}
\usepackage{graphicx} 
\usepackage{indentfirst}

\parskip=\medskipamount

\newcommand {\cD}{{\cal D}}

\newcommand {\cF}{{\cal F}}

\newcommand {\cH}{{\cal H}}

\newcommand {\cL}{{\cal L}}
\newcommand {\cM}{{\cal M}}
\newcommand {\cN}{{\cal N}}

\newcommand {\cP}{{\cal P}}

\newcommand {\cR}{{\cal R}}

\newcommand {\cV}{{\cal V}}
\newcommand {\cW}{{\cal W}}

\def\a{\alpha}
\def \bi{\bibitem}

\def\b{\beta}

\def\d{\delta}
\def\e{\epsilon}
\def\f{\phi}
\def\g{\gamma}
\def\G{\Gamma}

\def\k{\kappa}
\def\l{\lambda}
\def\m{\mu}
\def\n{\nu}
\def\o{\omega}

\def\q{\theta}
\def\r{\rho}
\def\s{\sigma}
\def\t{\tau}

\def\x{\xi}
\def\z{\zeta}
\def\D{\Delta}
\def\F{\Phi}
\def\J{\Psi}
\def\L{\Lambda}
\def\O{\Omega}

\def\S{\Sigma}
\def\U{\Upsilon}
\def\X{\Xi}

\newcommand{\ad}{{\dot{\alpha}}}                           
\newcommand{\bd}{{\dot{\beta}}}                            
\newcommand{\ve}{\varepsilon}                            

\newcommand{\pa}{\partial}                           
\newcommand{\hf}{\frac12}

\newcommand{\vf}{\varphi}
\newcommand{\sect}[1]{\setcounter{equation}{0}\section{#1}}

\newcommand{\be}{\begin{equation}}
\newcommand{\ee}{\end{equation}}
\newcommand{\bea}{\begin{eqnarray}}
\newcommand{\eea}{\end{eqnarray}}
\newcommand{\non}{\nonumber}
\newcommand{\1}{\underline{1}}
\newcommand{\2}{\underline{2}}

\def\dt#1{{\buildrel {\hbox{\LARGE .}} \over {#1}}}    

\newcommand{\bm}[1]{\mbox{\boldmath$#1$}}

\def\double #1{#1{\hbox{\kern-2pt $#1$}}}

\begin{document}

\begin{titlepage}

\begin{flushright}
hep-th/0601177\\
January, 2006\\
\end{flushright}
\vspace{5mm}

\begin{center}
{\Large \bf  
On  compactified harmonic/projective superspace,\\
5D superconformal theories, and all that}
\end{center}

\begin{center}

{\large  
Sergei M. Kuzenko\footnote{{kuzenko@cyllene.uwa.edu.au}}
} \\
\vspace{5mm}

\footnotesize{
{\it School of Physics M013, The University of Western Australia\\
35 Stirling Highway, Crawley W.A. 6009, Australia}}  
~\\

\vspace{2mm}

\end{center}
\vspace{5mm}

\begin{abstract}
\baselineskip=14pt
\noindent
Within the supertwistor approach, we analyse 
the superconformal structure of 4D $\cN=2$ compactified
harmonic/projective superspace. In the case of 5D 
superconformal symmetry, we derive the superconformal Killing vectors
and related building blocks which emerge in the transformation laws 
of  primary superfields. 
Various off-shell superconformal  multiplets
are presented  both in 5D harmonic and projective superspaces, 
including the so-called tropical  (vector) multiplet and polar 
(hyper)multiplet. Families of superconformal actions are 
described  both in the 5D harmonic and projective superspace
settings. We also present examples of 5D superconformal theories
with gauged central charge.
\end{abstract}

\vfill
\end{titlepage}

\newpage
\setcounter{page}{1}
\renewcommand{\thefootnote}{\arabic{footnote}}
\setcounter{footnote}{0}

\sect{Introduction}

According to Nahm's classification \cite{Nahm}, 
superconformal algebras exist in space-time dimensions 
${\rm D} \leq 6$.  Among the dimensions included,
the case of ${\rm D}=5$ is truly 
exceptional, for it allows  the existence of the unique superconformal algebra
$F(4)$ \cite{Kac}.
This  is in drastic  contrast to
the other dimensions   
which are known to be compatible with series of superconformal algebras 
(say,  4D $\cN$-extended or 6D $(\cN,0) $ superconformal  symmetry).
Even on formal grounds, the exceptional feature 
of the five-dimensional case 
is interesting enough
for studying in some depth the properties of 5D superconformal
theories. On the other hand, such  rigid superconformal  theories
are important  prerequisites in the  construction, 
within the superconformal tensor calculus 
\cite{Ohashi,Bergshoeff},
 of 5D supergravity-matter dynamical systems which 
are of primary importance, for example,  
in the context of bulk-plus-brane scenarios.

The main motivation for the present  work was the desire to develop 
a systematic setting to build 5D superconformal theories, and clearly this
is hardly possible without employing  superspace techniques.  
The superconformal algebra in five dimensions 
includes 5D simple (or $\cN=1$) 
supersymmetry algebra\footnote{On historic grounds,
5D simple supersymmetry is often labeled $\cN=2$,
see e.g. \cite{Bergshoeff}.}
as its super Poincar\'e subalgebra. 
As is well-known,  for supersymmetric theories 
in various dimensions with eight supercharges
(including the three important cases:
(i) 4D $\cN=2$, (ii) 5D $\cN=1$ and (iii) 6D $\cN= (1,0)$) 
a powerful formalism to generate off-shell formulations is the {\it harmonic 
superspace} approach that  was originally developed 
for the 4D $\cN=2$ supersymmertic Yang-Mills theories
 and supergravity \cite{GIKOS,GIOS}.
There also exists a somewhat different, but  related, formalism --
the so-called {\it projective superspace} approach 
\cite{projective0,Siegel,projective,BS}, first introduced
soon  after the harmonic superspace had appeared.
Developed originally for describing the general self-couplings
of 4D $\cN=2$ tensor multiplets, this approach has  been extended
to include some other interesting multiplets. 
Both the harmonic and projective approaches make use of 
the same superspace,  ${\mathbb R}^{D|8} \times S^2$, 
which first emerged, for $D=4$, in a work of Rosly \cite{Rosly}
(see also \cite{RS})
who built on earlier  ideas due to Witten \cite{Witten}.
In harmonic superspace,  one deals with  so-called Grassmann analytic
superfields required to be  smooth tensor fields on $S^2$.
In projective  superspace,  one also deals with  Grassmann analytic
superfields required, however, to be holomorphic on an open subset of $S^2$
(typically, the latter is chosen to be 
$ {\mathbb C}^* = {\mathbb C} \setminus \{ 0 \}$ 
in the Riemann sphere realisation 
$S^2 ={\mathbb C} \cup \{ \infty \}$).
In many respects, the harmonic and projective superspaces
are equivalent and complementary to each other \cite{Kuzenko}, 
although harmonic superspace is obviously more fundamental. 
Keeping in mind potential applications to the brane-world physics,
the projective superspace setting seems to be more useful, since the 5D projective 
supermultiplets \cite{KL} are easy to reduce to 4D $\cN=1$ superfields.

To our knowledge, no comprehensive discussion 
of the superconformal group 
and superconformal multiplets 
in projective superspace has been given, 
apart from  the analysis of $SU(2)$ invariance in \cite{projective0}
and  the semi-component consideration 
of tensor multiplets in \cite{dWRV}.
On the contrary, a realisation of 
the superconformal symmetry 
in 4D $\cN=2$ harmonic superspace\footnote{See
\cite{ISZ} for an extension to six dimensions.} 
has been known for 
almost twenty years \cite{GIOS-conf,GIOS}.
But some nuances of this realisation still 
appear to be quite mysterious 
(at least to newcomers) and call for a different interpretation.
Specifically,  one deals with 
superfields depending on harmonic variables $u^\pm_i$
subject to the two constraints 
\be
u^{+i}\,u^-_i =1~, \qquad  \overline{u^{+i}} =u^-_i~, \qquad \quad
i=\1, \2~
\label{1+2const}
\ee
when describing  general  4D $\cN=2$ super Poincar\'e 
invariant theories in harmonic superspace
\cite{GIKOS,GIOS}.
In the case of superconformal theories, on the other hand, 
only the first constraint in (\ref{1+2const}) has to be imposed
\cite{GIOS-conf,GIOS}.
Since any superconformal theory is, at the same time, 
a super  Poincar\'e  invariant one,  
some consistency  issues
seem to arise, such as that 
 of the functional spaces used to describe the theory.
It is quite remarkable that these issues simply do not occur  
if one pursues the twistor approach to  harmonic superspace 
\cite{Rosly2,LN,HH} (inspired by earlier constructions 
due to Manin  \cite{Manin}).
In such a setting, the constraints (\ref{1+2const})
can be shown to
appear  only as possible `gauge conditions'  
and therefore  they have no intrinsic significance, 
with the only structural condition being 
$u^{+i}\,u^-_i \neq 0$.
In our opinion, the supertwistor construction sketched 
in  \cite{Rosly2,LN}
and further analysed in \cite{HH}  is quite illuminating, 
for it allows a unified treatment 
of the harmonic and projective superspace formalisms.
That is why 
it is  reviewed  and developed further
in the present paper.
Unlike \cite{Rosly2,HH} and  standard texts on Penrose's twistor theory 
\cite{twistors}, see e.g. \cite{WW}, we avoid considering
compactified complexified Minkowski space and its 
super-extensions, for these concepts  are not relevant 
 from the point of view of 
superconformal model-building we are interested in.
Our 4D consideration is directly based on the use of 
(conformally) compactified Minkowski space $S^1 \times S^3$ 
and its super-extensions.
Compactified Minkowski space is quite interesting in its own right
(see, e.g. \cite{GS}), and its universal covering space
(i) possesses a unique causal structure compatible 
with conformal symmetry \cite{S}, 
and (ii) coincides with the boundary of five-dimensional 
Anti-de Sitter space, which is crucial in the context 
of the AdS/CFT duality   \cite{AGMOO}.

In the case of 5D superconformal symmetry, one can also pursue 
a supertwistor approach. However, since we are aiming at 
future applications to brane-world physics, 
a more pragmatic course is chosen here, 
which is based on the introduction of
the relevant superconformal Killing vectors
and  elaborating  associated building blocks. 
The concept of superconformal Killing  vectors
\cite{Sohnius,Lang,BPT,Shizuya,BK,HH,West}, has proved  
to be extremely useful for various studies of 
superconformal theories in four and six 
dimensions, see e.g.  \cite{Osborn,Park,KT}.

This paper is organized as follows.
In section 2 we review the construction \cite{U}
of  compactified 
Minkowski space $\overline{\cM}{}^4 = S^1 \times S^3$ 
as the set of null two-planes 
in the  twistor space. In section 
3 we discuss  $\cN$-extended compactified 
Minkowski superspace $\overline{\cM}{}^{4|4\cN}$
and introduce the corresponding superconformal 
Killing vectors. In section 4 we develop different
aspects of  4D $\cN=2$ compactified 
harmonic/projective superspace. 
Section 5 is devoted to the 5D superconformal 
formalism. Here we introduce the 5D superconformal 
Killing vectors and related building blocks, 
and also  introduce several off-shell superconformal
multiplets, both in the harmonic and projective superspaces.
Section 6 introduces the zoo of 5D superconformal theories.
Three technical appendices are also included at the end of the paper. 
In appendix  A, a non-standard realisation for $S^2$ is given.
Appendix B is devoted to the projective 
superspace action according to \cite{Siegel}.
Some aspects of the reduction \cite{KL}
from 5D projective supermultiplets to 4D $\cN=1,2$ 
superfields are collected in Appendix C.

\sect{Compactified Minkowski space}
\label{section:two}

We start by  recalling  a remarkable 
realisation\footnote{This realisation 
is known in the physics literature since the early 1960's
\cite{U,S,Tod,GS}, and it 
can be related (see, e.g. \cite{S})
to the Weyl-Dirac construction \cite{W,D}
of  compactified Minkowski space 
$ S^1 \times S^3 / {\mathbb Z}_2$
as the set of straight lines through the origin of the cone in
${\mathbb R}^{4,2}$.
In the mathematics literature, its roots go back 
to Cartan's classification of the irreducible homogeneous 
bounded symmetric domains \cite{Cartan,Hua}.} 
of  compactified 
Minkowski space $\overline{\cM}{}^4 = S^1 \times S^3$ 
as the set of null two-dimensional subspaces 
in the  twistor space\footnote{In the literature, 
the term `twistor space' is often used for ${\mathbb C}P^3$. 
In this paper we stick to the original 
Penrose terminonology \cite{twistors}.} 
which is a copy of 
${\mathbb C}^4$.
The twistor space is defined to be equipped with the scalar product
\bea
\langle T, S \rangle = T^\dagger \,\O \, S~, \qquad
\O =\left(
\begin{array}{cc}
 {\bf 1}_2  & 0\\
0 &      -{\bf 1}_2 
\end{array}
\right) ~,
\eea
for any twistors $T,S \in {\mathbb C}^4$.
By construction, this scalar product is invariant under the action
of the group $SU(2,2) $ 
to be identified with the conformal group.
The elements of $SU(2,2)$ will be represented by block 
matrices
\bea
g=\left(
\begin{array}{cc}
 A  & B\\
C &    D 
\end{array}
\right) \in SL(4,{\mathbb C}) ~, \qquad 
g^\dagger \,\O \,g = \O~,
\label{SU(2,2)}
\eea
where  $A,B,C$ and $D$ are $2\times 2$ matrices.

We will denote by $\overline{\cM}{}^4 $ the space of null 
two-planes through the origin 
in ${\mathbb C}^4$. Given a two-plane, it is generated 
by two linearly independent twistors $T^\m$, with $\m=1,2$,
such that
\be
\langle T^\m, T^\n \rangle = 0~, \qquad 
\m, \n =1,2~.
\label{nullplane1}
\ee 
Obviously, the basis  chosen, $\{T^\m\}$, is defined only modulo 
the equivalence relation 
\be
\{ T^\m \}~ \sim ~ \{ \tilde{T}^\m \} ~, \qquad
\tilde{T}^\m = T^\n\,R_\n{}^\m~, 
\qquad R \in GL(2,{\mathbb C}) ~.
\label{nullplane2}
\ee
Equivalently, 
the space  $\overline{\cM}{}^4 $ consists of 
$4\times 2$  matrices of rank two, 
\bea
( T^\m )=\left(
\begin{array}{c}
 F\\  G
\end{array}
\right) ~, \qquad
F^\dagger \,F =G^\dagger \,G~,
\label{two-plane}
\eea
where $F$ and $G$ are $2\times 2$ matrices  
defined modulo the equivalence relation 
\bea
\left(
\begin{array}{c}
 F\\  G
\end{array}
\right) ~ \sim ~
\left(
\begin{array}{c}
 F\, R\\  G\,R
\end{array}
\right) ~, \qquad R \in GL(2,{\mathbb C}) ~.
\eea

In order for 
the two twistors 
$T^\m $ in (\ref{two-plane})
to generate a two-plane, the $2\times 2$ matrices $F$ and $G$ 
must be non-singular, 
\be 
\det F \neq 0~, \qquad \det G \neq 0~.
\label{non-singular}
\ee
Indeed, let us suppose the opposite. 
Then, the non-negative 
Hermitian matrix $F^\dagger F$ has a zero eigenvalue.
Applying an equivalence transformation of the form
\bea
\left(
\begin{array}{c}
 F\\  G
\end{array}
\right) ~ \to ~
\left(
\begin{array}{c}
 F\, \cV\\  G\,\cV
\end{array}
\right) ~, \qquad \cV \in U(2) ~,
\non
\eea
and therefore 
\bea
F^\dagger \,F ~\to~ \cV^{-1} \Big(F^\dagger \,F \Big) \,\cV~, \qquad 
G^\dagger \,G ~\to~ \cV^{-1} \Big(G^\dagger \,G \Big)\, \cV~,
\non
\eea
we can arrive at the following situation 
\bea
F^\dagger \,F =G^\dagger \,G =
\left(
\begin{array}{cc}
 0  & 0\\
0 &    \l^2 
\end{array}
\right)~, \qquad  \l \in {\mathbb R} ~.
\non
\eea
In terms of the twistors $T^\m$,
the conditions obtained imply that $T^1 =0$ and $T^2 \neq 0$.
But this contradicts  the assumption that the two vectors $T^\m $ generate a two-plane.

Because of (\ref{non-singular}), we have 
\bea
\left(
\begin{array}{l}
 F\\  G
\end{array}
\right) ~ \sim ~
\left(
\begin{array}{c}
 h \\  {\bf 1}
\end{array}
\right) ~, \qquad h =F\,G^{-1}  \in U(2) ~.
\eea
It is seen  that the space $\overline{\cM}{}^4 $ 
can be identified with the group manifold 
$U(2) = S^1 \times S^3$.

The conformal group acts by linear transformations on the twistor space:
associated with the group element (\ref{SU(2,2)}) is
the transformation $T \to g\, T$, for any twistor $T \in {\mathbb C}^4$.
This group representation 
induces an action of $SU(2,2)$ on  $\overline{\cM}{}^4 $.
It is defined as follows:
\be
h ~\to ~g\cdot h = (A\,h +B ) \,(C\,h +D)^{-1} \in U(2) ~.
\ee

One can readily see  that $\overline{\cM}{}^4 $ is a homogeneous 
space of the group $SU(2,2)$, and therefore
it can be represented as $\overline{\cM}{}^4 =SU(2,2) /H_{h_0}$,
where $H_{h_0} $ is the isotropy group at a fixed unitary matrix 
$h_0 \in \overline{\cM}{}^4 $. With the choice 
\be
h_0 = - {\bf 1}~,
\ee
a coset representative $s(h)\in SU(2,2)$ that 
maps $h_0$ to $h \in \overline{\cM}{}^4 $ 
can be chosen as follows
(see, e.g. \cite{PS}):
\bea
s(h)=  (\det h)^{-1/4} 
\left(
\begin{array}{cr}
-h  ~ & 0  \\
0 ~&    {\bf 1}
\end{array}
\right)~, \qquad  
s(h) \cdot h_0 =h\in U(2)~.
\eea
The 
isotropy group corresponding to $h_0$
consists of those
$SU(2,2)$ group elements  (\ref{SU(2,2)}) which obey 
the requirement
\be
A+C = B+D~.
\label{stability}
\ee
This subgroup proves to be isomorphic to 
a group generated by the Lorentz transformations, dilatations and 
special conformal transformations. 
To visualise this,
it is useful to
implement a special similarity transformation for both the group 
$SU(2,2)$ and  the twistor space.

We introduce a special $4\times 4$ matrix $\S$,
\bea
\S= \frac{1}{  \sqrt{2} }
\left(
\begin{array}{cr}
 {\bf 1}_2  ~ & - {\bf 1}_2\\
{\bf 1}_2 ~&    {\bf 1}_2
\end{array}
\right)~, \qquad \S^\dagger \,\S= {\bf 1}_4~,
\eea
and associate with it the following similarity transformation:
\bea
g ~& \to & ~ {\bm g} = \S \, g\, \S^{-1} ~, \quad g \in SU(2,2)~; 
\qquad 
T ~ \to  ~ {\bm T} = \S \, T~, \quad T \in {\mathbb C}^4~.
\eea
The elements of $SU(2,2)$ are now represented by block 
matrices
\bea
{\bm g}=\left(
\begin{array}{cc}
 {\bm A}  & {\bm B}\\
{\bm C} &    {\bm D} 
\end{array}
\right) \in SL(4,{\mathbb C}) ~, \qquad 
{\bm g}^\dagger \,{\bm \O} \, {\bm g} = {\bm \O}~,
\label{SU(2,2)-2}
\eea
where 
\bea
{\bm \O} =  \S \, \O\, \S^{-1} 
=
\left(
\begin{array}{cc}
0&  {\bf 1}_2 \\
{\bf 1}_2 &0
\end{array}
\right) ~.
\eea
The $2\times 2$ matrices in (\ref{SU(2,2)-2}) are related to those
in (\ref{SU(2,2)})  as follows:
\bea
{\bm A} &=& \hf ( A+D-B-C)~, \non \\
{\bm B} &=& \hf ( A+B- C-D)~, \non \\
{\bm C} &=& \hf ( A+C-B-D)~, \non \\
{\bm D} &=& \hf ( A+B+C+D)~.
\eea
Now,  by comparing these expressions with (\ref{stability}) it is seen
that the stability group $\S H_{h_0} \S^{-1}$ 
consists of upper block-triangular matrices,
\be
{\bm C} =0~.
\label{stability2}
\ee

When applied to $\overline{\cM}{}^4 $, the effect 
of the similarity transformation\footnote{We follow the  two-component spinor 
notation of Wess and Bagger \cite{WB}.}
 is 
\bea
\left(
\begin{array}{c}
 h \\  {\bf 1}
\end{array}
\right) ~\to ~
\S\, \left(
\begin{array}{c}
 h \\  {\bf 1}
\end{array}
\right)  = \frac{1 }{ \sqrt{2} }
\left(
\begin{array}{c}
 h -{\bf 1} \\  h+{\bf 1}
\end{array}
\right) ~\sim ~
\left(
\begin{array}{c}
 {\bf 1} \\  -{\rm i}\, \tilde{x}
\end{array}
\right) ~,
\qquad  \tilde{x}  
=x^m \,(\tilde{\s}_m)^{\dt \a \a}~,
\label{two-plane-mod}
\eea
where 
\bea
-{\rm i}\, \tilde{x} = \frac{  h+{\bf 1} }{ h-{\bf 1} }~,
\qquad \tilde{x}^\dagger 
= \tilde{x}  ~.
\label{inverseCayley}
\eea
The inverse expression for $h$ in terms of $\tilde{x}$ 
is given by the so-called Cayley transform:
\be
-h = \frac{ {\bf 1} - {\rm i}\, \tilde{x} } { {\bf 1} + {\rm i}\, \tilde{x} } ~.
\label{Cayley}
\ee
It is seen that
\be
h_0 =-{\bf 1} \quad \longleftrightarrow \quad 
\tilde{x}_0 =0~.
\ee
Unlike the original twistor representation,
 eqs. (\ref{two-plane}) and (\ref{non-singular}),
the $2\times 2$ matrices  $h\pm{\bf 1}$ 
in (\ref{two-plane-mod}) may be singular at some points.
This means that the variables $x^m $ (\ref{inverseCayley}) 
are well-defined local coordinates in the open subset of $\overline{\cM}{}^4$
which is specified by $\det \,(  h-{\bf 1} ) \neq 0$ and, as will become clear soon, 
can be identified with the ordinary Minkowski space.

As follows from (\ref{two-plane-mod}), in terms of the 
variables $x^m$
the conformal group acts by fractional linear transformations 
\be
-{\rm i}\, \tilde{x}  ~\to ~-{\rm i}\, \tilde{x}' = 
\Big({\bm C} - {\rm i}\, {\bm D}\,\tilde{x}  \Big)
\Big({\bm A} - {\rm i}\, {\bm B}\,\tilde{x}  \Big)^{-1}~.
\ee
These transformations can be brought to a more familiar form 
if one takes into account the explicit structure of the 
elements of $SU(2,2)$:
\bea 
{\bm g} = {\rm e}^{\bm L}~, \quad 
{\bm L} = \left(
\begin{array}{cc}
\o_\a{}^\b - \hf \,\t \d_\a{}^\b  \quad &  -{\rm i} \,b_{\a \dt \b} 
\\
-{\rm i} \,a^{\dt \a \b} \quad & -{\bar \o}^{\dt \a}{}_{\dt \b} 
+ \hf  \, \t  \d^{\dt \a}{}_{\dt \b}   \\
\end{array}
\right)~,
\quad
{\bm L}^\dagger  =- {\bm \O} \, {\bm L} \, {\bm \O}~.
\label{confmat}
\eea
Here the matrix elements correspond to a
Lorentz transformation $(\o_\a{}^\b,~{\bar \o}^{\dt \a}{}_{\dt \b})$,
translation $a^{\dt \a \b}$, special conformal transformation
$ b_{\a \dt \b}$ and   dilatation $\t$. 
In accordance with (\ref{stability2}), the isotropy  group at $x_0=0$
is spanned by the Lorentz transformations, 
special conformal boosts and scale transformations.

\sect{Compactified Minkowski superspace}
\label{section:three}

The construction reviewed in the previous section 
can be immediately generalised to the case of 
$\cN$-extended  conformal supersymmetry 
 \cite{Manin},
by making  use of the  supertwistor 
space ${\mathbb C}^{4|\cN}$ introduced by Ferber \cite{Ferber}, 
with $\cN=1,2,3$ (the case $\cN=4$ is known to be somewhat special, 
and will not be discussed here). 
The supertwistor space is equipped with scalar product
\bea
\langle T, S \rangle = T^\dagger \,\O \, S~, \qquad
\O =\left(
\begin{array}{ccc}
 {\bf 1}_2  & {}&0 \\
{} &      -{\bf 1}_2 & {}\\
0 & {} & -{\bf 1}_{\cN} 
\end{array}
\right) ~,
\eea
for any supertwistors $T,S \in {\mathbb C}^{4|\cN}$.
The $\cN$-extended superconformal group acting on the supertwistor 
space is $SU(2,2|\cN) $. It is spanned by supermatrices of the form 
\bea
g \in SL(4|\cN ) ~, \qquad 
g^\dagger \,\O \,g = \O~.
\label{SU(2,2|N)}
\eea

In complete analogy with the bosonic construction, 
compactified Minkowski superspace $\overline{\cM}{}^{4|4\cN}$
is defined to be the space of null two-planes 
through the origin in  ${\mathbb C}^{4|\cN}$.
Given such a two-plane, it is generated by two  supertwistors 
$T^\m$ such that (i) their bodies are linearly independent;
(ii) they obey the equations
(\ref{nullplane1}) and (\ref{nullplane2}).  
Equivalently, 
the space  $\overline{\cM}^{4|4\cN} $ consists of 
rank-two supermatrices of the form 
\bea
( T^\m )=\left(
\begin{array}{c}
 F\\  G \\  
 \U
\end{array}
\right) ~, \qquad
F^\dagger \,F =G^\dagger \,G +\U^\dagger \,\U~,
\label{super-two-plane}
\eea
defined modulo the equivalence relation 
\bea
\left(
\begin{array}{c}
 F\\  G \\ \U
\end{array}
\right) ~ \sim ~
\left(
\begin{array}{c}
 F\, R\\  G\,R \\ \U\,R
\end{array}
\right) ~, \qquad R \in GL(2,{\mathbb C}) ~.
\eea
Here $F$ and $G$ are $2\times 2$ 
bosonic matrices, and $\U$ is a $\cN \times 2$ 
fermionic matrix.
As in  the bosonic case, we have
\bea
\left(
\begin{array}{c}
 F\\  G \\ \U
\end{array}
\right) ~ \sim ~
\left(
\begin{array}{c}
 h \\ {\bf 1}   \\ \Theta
\end{array}
\right) ~, \qquad 
h^\dagger h = {\bf 1} + \Theta^\dagger \, \Theta ~.
\eea

Introduce the supermatrix
\bea
\S= \frac{1 }{ \sqrt{2} }
\left(
\begin{array}{crc}
 {\bf 1}_2  ~ & - {\bf 1}_2 & 0\\
{\bf 1}_2 ~&    {\bf 1}_2 & 0  \\
0 & 0 ~& \sqrt{2} \,{\bf 1}_{\cN} 
\end{array}
\right)~, \qquad \S^\dagger \,\S= {\bf 1}_{4+\cN}~,
\eea
and associate with it the following similarity transformation:
\bea
g ~& \to & ~ {\bm g} = \S \, g\, \S^{-1} ~, \quad g \in SU(2,2|\cN)~; 
\qquad 
T ~ \to  ~ {\bm T} = \S \, T~, \quad T \in {\mathbb C}^{4|\cN}~.
\label{sim2}
\eea
The supertwistor metric becomes
\bea
{\bm \O} =  \S \, \O\, \S^{-1} 
=
\left(
\begin{array}{ccc}
0&  {\bf 1}_2 &0\\
{\bf 1}_2 &0 &0\\
0 & 0& -{\bf 1}_{\cN}
\end{array}
\right) ~.
\eea
When implemented on the superspace $\overline{\cM}{}^{4|4\cN} $,  
the similarity transformation results in 
\bea
\left(
\begin{array}{c}
 h \\  {\bf 1} \\ \Theta
\end{array}
\right) ~\to ~
\S\, \left(
\begin{array}{c}
 h \\  {\bf 1} \\ \Theta
\end{array}
\right)  = \frac{1 }{ \sqrt{2} }
\left(
\begin{array}{c}
 h -{\bf 1} \\  h+{\bf 1} \\ \sqrt{2}\, \Theta 
\end{array}
\right) ~\sim ~
\left(
\begin{array}{c}
 {\bf 1} \\  -{\rm i}\, \tilde{x}_+
 \\ 2 \,\q
\end{array}
\right) 
= \left(
\begin{array}{r}
\d_\a{}^\b \\  -{\rm i}\, \tilde{x}_+^{\dt \a \b}
 \\ 2 \,\q_i{}^\b
\end{array}
\right) 
~,
\label{super-two-plane-mod}
\eea
where 
\bea
-{\rm i}\, \tilde{x}_+ = \frac{  h+{\bf 1} }{ h-{\bf 1} }~,
\qquad 
\sqrt{2} \, \q = \Theta \,( h-{\bf 1} )^{-1}~.
\eea
The bosonic $\tilde{x}_+$ and fermionic $\q$ variables 
obey the reality condition
\be
\tilde{x}_+ -\tilde{x}_- =4{\rm i}\, \q^\dagger \,\q~,
\qquad \tilde{x}_- = (\tilde{x}_+)^\dagger~.
\label{chiral}
\ee
It is solved by 
\be 
x_\pm^{\dt \a \b} = x^{\dt \a \b} \pm 2{\rm i} \, 
{\bar \q}^{\dt \a i} \q^\b_i ~,\qquad 
{\bar \q}^{\dt \a i} = \overline{ \q^\a_i}~, 
\qquad \tilde{x}^\dagger = \tilde{x}~,
\ee
with $z^A = (x^a ,\q^\a_i , {\bar \q}_{\dt \a}^i)$ the coordinates
of $\cN$-extended flat global superspace ${\mathbb R}^{4|4\cN}$.
We therefore see that the supertwistors in 
(\ref{super-two-plane-mod}) are parametrized 
by the variables $x^a_+$ and $\q^\a_i$ which are 
the coordinates in the chiral subspace.
Since the superconformal group acts by linear
transformations on ${\mathbb C}^{4| 2\cN}$, 
we can immediately conclude that 
it acts by holomorphic transformations 
on the chiral subspace.

To describe the action of $SU(2,2|\cN)$ on the chiral subspace, 
let us consider a generic group element:
\be
{\bm g} ={\rm e}^{\bm L}~, \quad
{\bm L} = \left(
\begin{array}{ccc}
\o_\a{}^\b - \s \d_\a{}^\b  \quad &  -{\rm i} \,b_{\a \dt \b} \quad &
2\eta_\a{}^j \\
 -{\rm i} \,a^{\dt \a \b} \quad & -{\bar \o}^{\dt \a}{}_{\dt \b} 
+ {\bar \s}  \d^{\dt \a}{}_{\dt \b}   \quad &
2{\bar \e}^{\dt \a j} \\
2\e_i{}^\b \quad & 2{\bar \eta}_{i \dt \b} \quad & \frac{2}{\cN}({\bar \s} - \s)\,
\d_i{}^j +  \L_i{}^j
\end{array}
\right)~,
\label{su(2,2|n)}
\ee
where
\be
\s = \hf \left( \t + {\rm i}\, 
\frac{\cN}{\cN -4} \vf \right)~,
\qquad
\L^\dag =  -\L~, \qquad  {\rm tr}\; \L = 0~.
\ee
Here the matrix elements, which 
are not present in (\ref{confmat}),  correspond to 
a $Q$--supersymmetry $(\e_i^\a,~ {\bar \e}^{\dt \a i})$,
$S$--supersymmetry $(\eta_\a^i,~{\bar \eta}_{i \dt \a})$,
combined scale and chiral transformation $\s$, 
and chiral $SU(\cN)$ transformation $\L_i{}^j$.
Now, one can check that 
the coordinates of  the chiral subspace transform 
as follows:
\bea
\d \tilde{x}_+ &=& \tilde{a}  +(\s +{\bar \s})\, \tilde{x}_+
-{\bar \o}\, \tilde{x}_+ -\tilde{x}_+ \,\o 
+\tilde{x}_+ \,b \,\tilde{x}_+
+4{\rm i}\, {\bar \e} \, \q - 4 \tilde{x}_+ \, \eta \, \q ~,
\non \\
\d \q &=& \e + \frac{1}{\cN} \Big( 
(\cN-2) \s + 2 
{\bar \s}\Big)\, \q - \q\, \o 
+ \L \, \q  +\q \, b \,  \tilde{x}_+
-{\rm i}\,{\bar \eta}\, \tilde{x}_+ - 4\,\q \,\eta \, \q~.
\label{chiraltra}
\eea

Expressions (\ref{chiraltra})
can be rewritten in a more compact form,
\be
\d x^a_+ = \x^a_+ (x_+, \q) ~, \qquad
\d \q^\a_i = \x^\a_i (x_+, \q)  ~,
\ee
where 
\be
\x^a_+ = \x^a + \frac{\rm i}{8} \,\x_i \,\s^a \, {\bar \q}^i~, 
\qquad \overline{\x^a} =\x^a~.
\ee
Here the parameters $\x^a$ and $\x^\a_i$ are components
of the superconformal  Killing vector
\be
\x = {\overline \x} = \x^a (z) \,\pa_a + \x^\a_i (z)\,D^i_\a
+ {\bar \x}_{\dt \a}^i (z)\, {\bar D}^{\dt \a}_i~,
\ee   
which generates the infinitesimal transformation in the full superspace,
$z^A \to z^A + \x \,z^A$, and is
defined to satisfy 
\be
[\x \;,\; {\bar D}_i^\ad] \; \propto \; {\bar D}_j^\bd ~,
\ee   
and therefore
\be
{\bar D}_i^{\dt \a } \x^{\dt \b \b} = 4{\rm i} \, \ve^{\dt \a{}\dt \b} \,\x^\b_i~.
\label{4Dmaster}
\ee

All information about the superconformal algebra is encoded 
in the superconformal Killing vectors.
${}$From eq. (\ref{4Dmaster})  it follows that
\be
[\x \;,\; D^i_\a ] = - (D^i_\a \x^\b_j) D^j_\b
= {\tilde\o}_\a{}^\b  D^i_\b - \frac{1}{\cN}
\Big( (\cN-2) \tilde{\s} + 2 \overline{ \tilde{\s}}  \Big) D^i_\a
- \tilde{\L}_j{}^i \; D^j_\a \;.
\label{4Dmaster2} 
\ee
Here the parameters of `local' Lorentz $\tilde{\o}$ and
scale--chiral $\tilde{\s}$ transformations are
\be
\tilde{\o}_{\a \b}(z) = -\frac{1}{\cN}\;D^i_{(\a} \x_{\b)i}\;,
\qquad \tilde{\s} (z) = \frac{1}{\cN (\cN - 4)}
\left( \hf (\cN-2) D^i_\a \x^\a_i - 
{\bar D}^{\dt \a}_i {\bar \x}_{\dt \a}^{ i} \right)
\label{lor,weyl}
\ee
and turn out to be chiral
\be
{\bar D}_{\dt \a i} \tilde{\o}_{\a \b}~=~ 0\;,
\qquad {\bar D}_{\dt \a {} i} \tilde{\s} ~=~0\;.
\ee
The parameters $\tilde{\L}_j{}^i$ 
\be
\tilde{\L}_j{}^i (z) = -\frac{\rm i}{32}\left(
[D^i_\a\;,{\bar D}_{\dt \a j}] - \frac{1}{\cN}
\d_j{}^i  [D^k_\a\;,{\bar D}_{\dt \a k}] \right)\x^{\dt \a \a}~, \qquad
\tilde{\L}^\dag = - \tilde{\L}~, \qquad  {\rm tr}\; \tilde{\L} = 0
\label{lambda}
\ee
correspond to `local' $SU(\cN )$ transformations.
One can readily check the identity 
\be
D^k_\a \tilde{\L}_j{}^i = -2 \left( \d^k_j D^i_\a 
-\frac{1}{\cN} \d^i_j D^k_\a  \right) \tilde{\s}~.
\label{an1}
\ee

\sect{Compactified harmonic/projective superspace}
\label{section:four}

${}$For Ferber's  supertwistors used in the previous section, 
a more appropriate name seems to  be  {\it even supertwistors}.
Being elements of ${\mathbb C}^{4|\cN}$, these objects have 
four bosonic components and $\cN$ fermionic components.
One can also consider {\it odd supertwistors} \cite{LN}. By definition, 
these are $4+\cN$ vector-columns such that their top four entries
are fermionic,  and the rest $\cN$ components are bosonic. 
In other words, the odd supertwistors are elements of 
${\mathbb C}^{\cN |4}$. It is natural to treat the even and odd supertwistors
as the even and odd elements, respectively, 
of a supervector space\footnote{See, e.g.  \cite{DeWitt,BK}
for reviews on supervector spaces.}
of dimension $(4|\cN )$ 
on which the superconformal group $SU(2,2|\cN) $ acts.
Both even and odd supertwistors should be used \cite{Rosly2,LN} in order 
to define harmonic-like superspaces in extended supersymmetry.

Throughout this section, our consideration is restricted to the case  $\cN=2$.
Then, $\tilde{\L}^{ij} = \ve^{ik} \,\tilde{\L}_k{}^{j}$ is symmetric, 
$\tilde{\L}^{ij}= \tilde{\L}^{ji}$,
and eq. (\ref{an1})  implies
\be
D^{(i}_\a \tilde{\L}^{jk)} = {\bar D}^{(i}_{\dt \a} \tilde{\L}^{jk)}= 0~.
\label{an2}
\ee

\subsection{Projective realisation}

${}$ Following   \cite{LN}, 
we accompany the two even null supertwistors $T^\m$, 
which occur in the construction of  the compactified
$\cN=2 $ superspace $\overline{\cM}{}^{4|8} $,
by an odd supertwistor $\X$ with non-vanishing  {\it body}
(in particular, the body of $ \langle \X, \X \rangle$ is non-zero).
These supertwistors are required to obey 
\be
\langle T^\m, T^\n \rangle = \langle T^\m, \X \rangle = 
0~, \qquad 
\m, \n =1,2 ~,
\label{nullplane3}
\ee 
and are defined modulo the equivalence  relation
\bea
(\X, T^\m)~\sim ~  (\X, T^\n) \,
\left(
\begin{array}{cc}
 c~  &0  \\  
 \r_\n~ & R_\n{}^\m 
\end{array}
\right) ~,\qquad 
\left(
\begin{array}{cc}
 c~  &0  \\  
 \r~ & R
\end{array}
\right) \in GL(1|2)~,
\eea
with $\r_\n$  anticommuting complex parameters.
The superspace obtained can be seen to be 
$\overline{\cM}{}^{4|8} \times S^2$.
Indeed, using the above freedom in the definition 
of $T^\m$ and $\X$, we can choose them to be of the form
\bea
T^\m \sim  
\left(
\begin{array}{c}
 h \\ {\bf 1}   \\ \Theta
\end{array}
\right) ~, 
\qquad 
\X \sim
\left(
\begin{array}{c}
0    \\ - \Theta^\dagger \,v \\ v
\end{array}
\right) ~, 
\qquad 
h^\dagger h = {\bf 1} + \Theta^\dagger \, \Theta ~, 
\quad v \neq 0~.
\eea
Here 
the non-zero two-vector $v \in {\mathbb C}^2$ is still defined
modulo re-scalings $v \to c\, v $, with $c \in {\mathbb C}^*$.
A natural name for the supermanifold obtained is 
{\it projective superspace}.

\subsection{Harmonic realisation}
Now, we would like to present a somewhat  different, but equivalent, 
realisation for $\overline{\cM}{}^{4|8} \times S^2$
inspired by the exotic 
realisation for the two-sphere described in Appendix A.
We will consider a space of quadruples $\{T^\m, \X^+, \X^- \}$
consisting of two even supertwistors $T^\m$ and 
two odd supertwistors $\X^\pm$ such that (i) the bodies of 
$T^\m$ are linearly independent four-vectors; 
(ii) the bodies of $\X^\pm$ are lineraly independent two-vectors.
These supertwistors
are further required to obey the relations 
\be
\langle T^\m, T^\n \rangle = \langle T^\m, \X^+ \rangle = 
\langle T^\m, \X^- \rangle = 
0~, \qquad 
\m, \n =1,2 ~,
\label{nullplane4}
\ee 
and are defined modulo the equivalence  relation
\bea
(\X^-,\X^+, T^\m)\sim   (\X^-,\X^+, T^\n) \,
\left(
\begin{array}{ccc}
a~& 0~& 0 \\
b~& c~  &0  \\  
 \r^-_\n~ & \r^+_\n ~&R_\n{}^\m 
\end{array}
\right) ~,\quad 
\left(
\begin{array}{lll}
a~& 0~& 0 \\
b~& c~  &0  \\  
 \r^- ~ & \r^+  ~&R 
\end{array}
\right)
 \in GL(2|2)~,
\eea
with $\r^\pm_\n$  anticommuting complex parameters.
Using the `gauge freedom' in the definition 
of $T^\m$ and $\X^\pm$, these supertwistors
 can be chosen to have  the form
\bea
T^\m \sim  
\left(
\begin{array}{c}
 h \\ {\bf 1}   \\ \Theta
\end{array}
\right) ~, 
\quad 
\X^\pm \sim
\left(
\begin{array}{c}
0    \\ - \Theta^\dagger \,v^\pm \\ v^\pm
\end{array}
\right) ~, 
\quad 
h^\dagger h = {\bf 1} + \Theta^\dagger \, \Theta ~, 
\quad
\det \, (v^- \,v^+) \neq 0~.
\eea
Here the `complex harmonics'  $v^\pm$
are still defined modulo  
arbitrary transformations of the form (\ref{equivalence2}).
Given a $2\times 2$   matrix ${\bm v}=
(v^-\, v^+ ) \in  GL(2,{\mathbb C})$, 
there always exists a lower triangular matrix $R$ such that 
${\bm v} R \in SU(2)$. The latter  implies that $v^-$ is uniquely 
determined in terms of  $v^+$, and therefore 
the supermanifold under consideration is indeed
$\overline{\cM}{}^{4|8} \times S^2$.
In accordance with the construction given, 
a natural name for this supermanifold   is 
{\it harmonic superspace}.

\subsection{Embedding of 
$\bm{ {\mathbb R}^{4|8} \times S^2}$:
Harmonic realisation}

We can now analyse the structure of superconformal 
transformations on the flat global superspace
$ {\mathbb R}^{4|8} \times S^2$
embedded in   $\overline{\cM}{}^{4|8} \times S^2$.

Upon implementing the similarity transformation, eq. (\ref{sim2}),
we have
\bea
({\bm T}^\m ) \sim 
\left(
\begin{array}{c}
 {\bf 1} \\  -{\rm i}\, \tilde{x}_+
 \\ 2 \,\q
\end{array}
\right) 
= \left(
\begin{array}{r}
\d_\a{}^\b \\  -{\rm i}\, \tilde{x}_+^{\dt \a \b}
 \\ 2 \,\q_i{}^\b
\end{array}
\right) 
~, \qquad 
 {\bm \X}^\pm \sim  
 \left(
\begin{array}{c}
 0 \\   2{\bar \q}^\pm 
 \\  u^\pm
\end{array}
\right) 
=  \left(
\begin{array}{c}
 0 \\   2{\bar \q}^{\pm \dt \a }
 \\  u^\pm_i
\end{array}
\right) ~.
\label{par}
\eea
with 
\bea
\det  \Big(u_i{}^- \, u_i{}^+ \Big) =
 u^{+i} \,u^-_i \neq 0~,
\qquad u^{+i} = \ve^{ij} \,u^+_j~.
\non
\eea
Here the bosonic $x^m_+$ and fermionic $\q^\a_i$ variables 
are  related to each other by the reality condition (\ref{chiral}).
The orthogonality conditions
$\langle {\bm T}^\m, {\bm \X}^\pm \rangle = 0$ imply
\be
{\bar \q}^{+ \dt \a } = {\bar \q}^{\dt \a i} \,u^+_i~,
\qquad 
{\bar \q}^{- \dt \a } = {\bar \q}^{\dt \a i} \,u^-_i~.
\ee
The complex harmonic variables $u^\pm_i$ 
in (\ref{par}) are still defined  modulo arbitrary
transformations of the form 
\bea
\Big(u_i{}^- \, u_i{}^+ \Big)  ~\to ~
\Big(u_i{}^- \, u_i{}^+ \Big) 
\,R~,
\qquad 
R= \left(
\begin{array}{cc}
 a  & 0\\
b &      c 
\end{array}
\right) \in GL(2,{\mathbb C})~.
\label{equivalence22}
\eea

The `gauge' freedom (\ref{equivalence22})
can be reduced by imposing the `gauge' condition
\be 
u^{+i} \,u^-_i =1~.
\label{unimod}
\ee
It can be further reduced by choosing 
 the harmonics to obey the reality condition
\be
u^{+i} =\overline{u^-_i} ~.
\label{real}
\ee
Both requirements (\ref{unimod}) and (\ref{real}) have no fundamental 
significance, and  represent themselves
possible  gauge conditions only.
It is worth pointing out that the reality condition 
(\ref{real}) implies
$ \langle {\bm \X}^- ,  {\bm \X}^+ \rangle = 0$.
If both equations (\ref{unimod}) and (\ref{real}) hold, 
then we have in addition 
$ \langle {\bm \X}^+ ,  {\bm \X}^+ \rangle 
=  \langle {\bm \X}^- ,  {\bm \X}^- \rangle = -1$.

In what follows, the harmonics will be assumed 
to obey  eq. (\ref{unimod}) only. 
As explained in the appendix, the gauge freedom
(\ref{equivalence22}) allows one to represent 
any infinitesimal transformation of the harmonics
as follows:
\bea
\d u^-_i =0~, \qquad \d u^+_i  = \r^{++}(u)\, u^-_i~, 
\non
\eea
for some parameter $\r^{++}$ which is determined by 
the transformation under consideration.

In the case of an infinitesimal superconformal transformation
 (\ref{su(2,2|n)}),
one derives
\bea
\d u^-_i =0~, \qquad 
\d u^+_i  = - \tilde{\L}^{++}\, u^-_i~,
\qquad 
\tilde{\L}^{++} = \tilde{\L}^{ij} \,u^+_i u^+_j~,
\label{deltau+}
\eea
with the parameter $ \tilde{\L}^{ij} $ given by (\ref{lambda}).
Due to (\ref{an2}), we have (using the notation $D^\pm_\a =D^i_\a u^\pm_i$ 
and $ {\bar D}^\pm_{\dt \a} ={\bar D}^i_{\dt \a} u^\pm_i$) 
\be
D^+_\a  \tilde{\L}^{++} ={\bar D}^+_{\dt \a}  \tilde{\L}^{++} =0~,
\qquad D^{++}  \tilde{\L}^{++} =0~.
\label{L-anal}
\ee
Here and below, we make use of the harmonic derivatives \cite{GIKOS}
\bea
D^{++}=u^{+i}\frac{\partial}{\partial u^{- i}} ~,\qquad
D^{--}=u^{- i}\frac{\partial}{\partial u^{+ i}} ~.
\label{5}
\eea
It is not difficult to express $\tilde{\L}^{++} $ in terms
of the parameters in (\ref{su(2,2|n)}) and superspace coordinates:
\be
\tilde{\L}^{++} =\L^{ij} \,u^+_i u^+_j +4 \, {\rm i}\,\q^+ \,b \,{\bar \q}^+
- ( \q^+ \eta^+ -{\bar \q}^+ {\bar \eta}^+ ) ~.
\ee 
The transformation (\ref{deltau+}) coincides with the one 
originally given in 
\cite{GIOS-conf}.

${}$For the superconformal variations
of $\q^{+}_{ \a} $ and ${\bar \q}^+_{\dt \a}$, one finds
\bea
\d \q^{+}_{ \a} &=& \d \q^i_{  \a  } \, u^+_i + \q^i_{  \a  }\,  \d u^+_i 
= \x^i_{  \a  } \, u^+_i -
 \tilde{\L}^{++} \, \q^i_{  \a  } \, u^-_i ~,
\eea
and similarly for $\d {\bar \q}^{+}_{\dt \a}$. From eqs. (\ref{4Dmaster2}) 
and (\ref{L-anal}) one then deduces 
\be
D^+_\b \, \d q^{+}_{ \a } = {\bar D}^+_{\dt \b} \,  \d \q^{+}_{ \a}=0~,
\ee
and similarly for $\d {\bar \q}^{+}_{\dt \a}$.
The superconformal variations $ \d q^{+}_{ \a } $ and 
$\d {\bar \q}^{+}_{ \dt \a}$ can be seen to coincide 
with those originally given in  \cite{GIOS-conf}.
One can also check that the superconformal variation of
the analytic bosonic coordinates
\be
y^a = x^a - 2{\rm i}\, \q^{(i}\s^a {\bar \q}^{j)}u^+_i u^-_j~,
\qquad 
D^+_\b \, y^a = {\bar D}^+_{\dt \b} \, y^a=0~, 
\ee
is analytic. This actually follows from the transformation
\be
\d D^+_\a \equiv
[ \x - \tilde{\L}^{++} D^{--}  ,   D^+_{ \a} ]
=  \tilde{\o}_{ \a}{}^{ \b}\, D_{ \b}^+
- ( \tilde{\s}  +   \tilde{\L}^{ij} \,u^+_i u^-_j  ) \, D^+_{ \a}~,
\ee
and similarly for $\d {\bar D}^+_{\dt \a} $.

We conclude that the analytic subspace parametrized by 
the variables 
$$\z=( y^a,\q^{+\a},{\bar\q}^+_{\dt \a}, \,
u^+_i,u^-_j )~,
\qquad D^+_\b \, \z = {\bar D}^+_{\dt \b} \, \z=0~, 
$$
is invariant under the superconformal group.
The superconformal variations of these coordinates
coincide with  those given in  \cite{GIOS-conf}.
No consistency clash occurs between 
the $SU(2)$-type constraints (\ref{1+2const}) 
and the superconformal transformation law (\ref{deltau+}),
because the construction does not require 
imposing either of the constraints (\ref{1+2const}).

Using eq. (\ref{an1}) one can show that 
the following descendant
of the superconformal Killing vector 
\be
\S = \tilde{\L}^{ij} \,u^+_i u^-_j + \tilde{\s} 
+\overline{ \tilde{\s} }
\ee
possesses the properties
\be
D^+_\b \, \S = {\bar D}^+_{\dt \b} \, \S=0~, \qquad 
D^{++} \S =\tilde{\L}^{++}~.
\ee
It turns out that the objects  $\x$, $\tilde{\L}^{++}$ and $\S$ determine
the superconformal transformations of primary analytic superfields
\cite{GIOS}.

\subsection{Embedding of $\bm{ {\mathbb R}^{4|8} \times S^2}$: 
Projective realisation}

Now, let us try to exploit 
the realisation of $S^2$ as the Riemann sphere ${\mathbb C}P^1$.
The superspace can be covered by two open sets -- the north chart 
and the south chart -- that are specified by the 
conditions: (i) $u^{+ \underline{1}} \neq 0$; and 
(ii)  $u^{+ \underline{2}} \neq 0$.

In the north chart, the gauge freedom (\ref{equivalence22}) 
can be completely fixed by choosing 
\bea
 u^{+i} \sim (1, w) \equiv w^i ~, \quad && \quad u^+_i \sim (-w,1) = w_i~, 
\quad \qquad \non \\
 u^{-i} \sim (0,-1) ~, \quad && \quad u^-_i \sim (1,0)~.
\label{projectivegaugeN}
\eea
Here $w$ is the complex coordinate parametrizing  the north chart. 
Then the  transformation law (\ref{deltau+}) turns into
\be
\d w =  \tilde{\L}^{++}(w)~,
\qquad 
\tilde{\L}^{++} (w)= \tilde{\L}^{ij} \,w^+_i w^+_j~.
\label{deltaw+}
\ee
It is seen that the superconformal group acts by holomorphic
transformations. 

The south chart is defined by 
\bea
u^{+i} \sim (y, 1) \equiv y^i~, \quad && \quad
u^+_i \sim (-1,y) =y_i ~,  \non \\
\quad \qquad 
u^{-i} \sim (1,0) ~, \quad  && \quad u^-_i \sim (0,1)~,
\eea
with $y$ the local complex coordinate. The  transformation law (\ref{deltau+}) 
becomes 
\be
\d y = -\tilde{\L}^{++}(y)~,
\qquad 
\tilde{\L}^{++} (y)= \tilde{\L}^{ij} \,y^+_i y^+_j~.
\label{deltay+}
\ee

In the overlap of the north and south  charts, 
the corresponding complex coordinates
are related to each other in the standard way:
\be
y= \frac{1}{w}~.
\ee

\sect{5D superconformal formalism}
\label{section:five}

As we have seen, modulo some global topological issues, 
all information about the superconformal structures 
in a superspace is encoded in the corresponding 
superconformal Killing vectors.
In developing the  5D superconformal formalism below, 
we will not pursue global aspects, and simply base 
our consideration upon elaborating 
the superconformal Killing vectors and related 
concepts.
Our 5D notation and conventions follow \cite{KL}.

\subsection{5D superconformal Killing vectors}

In 5D simple
superspace ${\mathbb R}^{5|8}$ parametrized  
by  coordinates  $ z^{\hat A} = (x^{\hat a},  \q^{\hat \a}_i )$, 
we introduce an infinitesimal coordinate transformation 
\be
 z^{\hat A} ~\to ~ z^{\hat A} =  z^{\hat A}  + \x \, z^{\hat A} 
\ee
generated by a real vector field
\be
\x ={\bar \x} = \x^{\hat a} (z) \, \pa_{\hat a} 
+  \x^{\hat \a}_i (z) \, D_{\hat \a}^i ~,
\ee
with $D_{\hat A} = ( \pa_{\hat a}, D_{\hat \a}^i ) $
the flat covariant derivatives.
The transformation is said to be superconformal if 
$[\x , D_{\hat \a}^i] \propto D_{\hat \b}^j $,
or more precisely 
\be
[\x , D_{\hat \a}^i] =  -( D_{\hat \a}^i \, \x^{\hat \b}_j )\, D_{\hat \b}^j~.
\label{master1}\ee
The latter equation is equivalent to 
\be 
D_{\hat \a}^i \x^{\hat b} = 2{\rm i} \,(\G^{\hat b})_{\hat \a}{}^{\hat \b}\,
\x^i_{\hat \b}
 = - 2{\rm i} \,(\G^{\hat b})_{\hat \a \hat \b}\,
\x^{\hat \b i} ~.
\label{master2}
\ee
It follows from here
\be
\ve^{ij} \,(\G_{\hat a})_{\hat \a \hat \b}\, \pa^{\hat a} \x^{\hat b}
= (\G^{\hat b})_{\hat \a \hat \g}\,    D_{\hat \b}^j \, \x^{\hat \g i} 
+ (\G^{\hat b})_{\hat \b \hat \g}\,  D_{\hat \a}^i \, \x^{\hat \g j}~.
\label{master3}
\ee
This equation implies that $\x^{\hat a}= \x^{\hat a}(x,\q)  $ is 
an ordinary  conformal Killing vector, 
\be
\pa^{\hat a} \x^{\hat b}+\pa^{\hat b} \x^{\hat a} 
=\frac{2}{5}\, \eta^{\hat a \hat b} \,
\pa_{\hat c} \, \x^{\hat c}~,
\label{master4}
\ee
depending parametrically on the Grassmann superspace coordinates, 
\bea 
 \x^{\hat a}(x,\q) &=& b^{\hat a} (\q) + 2\s (\q) \, x^{\hat a} 
+ \o^{\hat a}{}_{\hat b} (\q) \,x^{\hat b} 
+k^{\hat a} (\q)\, x^2 -2 x^{\hat a} x_{\hat b}\, k^{\hat b}(\q)  ~,
\eea
with $\o^{\hat a \hat b} =- \o^{\hat a \hat b}$.

${}$From (\ref{master2}) one can derive a closed equation
on the vector components 
$\x_{\hat \b \hat \g} = (\G^{\hat b})_{\hat \b \hat \g}   \x_{\hat b}$:  
\be
D^i_{( \hat \a}\, \x_{\hat \b  ) \hat \g} 
=-\frac{1}{5}  \,D^{ \hat \d  i} \, \x_{\hat \d ( \hat \a}  \, \ve_{\hat \b ) \hat \g}~.
\ee
One can also deduce closed equations on  the spinor
components $ \x^{\hat \a}_i $:
\bea
D_{\hat \a}^{(i} \, \x_{\hat \b}^{ j) } &=&\frac{1}{ 4} \,
\ve_{\hat \a \hat \b} \,D^{\hat \g (i }\,  \x_{\hat \g}^{ j)}~,  
\label{master5} \\
(\G^{\hat b})_{\hat \a \hat \b} \,D^{\hat \a i} \x^{\hat \b }_i
&=&0~.
\label{master6}
\eea

At this stage it is useful to let harmonics  
$u^\pm_i$, such that $u^{+i}u^-_i\neq 0$,
enter the scene  for the first  time.
With the definitions  
$D^\pm_{\hat \a} = D^i_{\hat \a} \, u^\pm_i$ and 
$\x^\pm_{\hat \a} = \x^i_{\hat \a} \, u^\pm_i$,
eq. (\ref{master5}) is equivalent  to
\be
D_{\hat \a}^{+} \x_{\hat \b}^{ + } =\frac{1}{4} \,
\ve_{\hat \a \hat \b} \,D^{+ \hat \g  } \x_{\hat \g}^{ +}
\quad \Longrightarrow \quad 
D_{\hat \a}^{+} D_{\hat \b}^{+} \x_{\hat \g}^{ + } =0~.
\ee

The above results lead to 
\be
[\x , D_{\hat \a}^i] = \tilde{\o}_{\hat \a}{}^{\hat \b}\, D_{\hat \b}^i
-\tilde{\s} \, D_{\hat \a}^i - \tilde{\L}_j{}^i D_{\hat \a}^j~,
\label{param}
\ee
where 
\bea
\tilde{\o}^{\hat \a \hat \b}  =-\hf \,D^{k (\hat \a} \x^{ \hat \b  )}_k~,
\quad
 \tilde{\s} = \frac{1}{8} D_{\hat \g}^k \x^{\hat \g }_k~,
\quad \tilde{\L}^{ij} = \frac{1}{ 4} 
D_{\hat \g }^{( i} \x^{j) \hat \g }~.
\eea

The parameters on the right of (\ref{param}) are related to each other
as follows
\bea
D_{\hat \a}^i \tilde{\o}_{\hat \b \hat \g} &=& 
2\Big( \ve_{\hat \a \hat \b} \, D_{\hat \g}^i \tilde{\s}
+ \ve_{\hat \a \hat \g} \, D_{\hat \b}^i \tilde{\s} \Big)~, \non \\
D_{\hat \a}^i \tilde{\L}^{jk} &=& 
3\Big( \e^{ik} \,D_{\hat \a}^j \tilde{\s}  +
 \e^{ij} \,D_{\hat \a}^k \tilde{\s} \Big) ~.
\label{relations}
\eea

The superconformal transformation of the 
superspace integration measure 
involves
\be
\pa_{\hat a} \,\x^{\hat a}  - D^i_{\hat \a} \,\x^{\hat \a}_i 
=2\tilde{\s}~.
\label{trmeasure1}
\ee

\subsection{Primary superfields}
Here we give a few examples of 5D primary superfields,
 without Lorentz indices.

Consider a completely symmetric iso-tensor  
superfield $H^{i_1\dots i_n}= H^{(i_1\dots i_n)}$ 
with the superconformal transformation law
\be
\d H^{i_1\dots i_n}=  -\x \,H^{i_1\dots i_n}
-p \,\tilde{\s}\, H^{i_1\dots i_n}
-\tilde{\L}_k{}^{(i_1} H^{i_2\dots i_n )k} ~,
\label{lin1}
\ee
with $p$ a constant parameter being equal to  half 
the conformal weight  of $H^{i_1\dots i_n}$. 
It turns out that this parameter is equal to
$3n$ if $H^{i_1\dots i_n}$ is 
constrained by
\be
D_{\hat \a}{}^{(j} H^{i_1\dots i_n)} =0 \quad 
\longrightarrow \quad p=3n~.
\label{lin2}
\ee

The vector multiplet strength transforms as
\be
\d W = - \x\,W -2\tilde{\s} \,W~.
\label{vmfstransfo}
\ee
The conformal weight of $W$ is uniquely fixed by 
the Bianchi identity
\be
D^{(i}_{\hat \a} D_{\hat \b }^{j)}  W
= \frac{1 }{ 4} \ve_{\hat \a \hat \b} \,
D^{\hat \g (i} D_{\hat \g }^{j)}  W~
\label{Bianchi1}
\ee
obeyed by $W$.

\subsection{Analytic building blocks}
In what follows we make use of the harmonics $u^\pm_i$
subject to eq. (\ref{unimod}). As in the 4D $\cN=2$ case,
eq. (\ref{unimod})
has no intrinsic significance, with the only essential condition being
$(u^+u^-) \equiv u^{+i}u^-_i\neq 0$. Eq.  (\ref{unimod}) is nevertheless
handy, for it allows one to get rid of numerous  annoying factors 
of $(u^+u^-)$.

Introduce
\be
\S = \tilde{\L}^{ij} \,u^+_i u^-_j +3\tilde{\s}~,\qquad 
\tilde{\L}^{++} = D^{++} \S
=\tilde{\L}^{ij} \,u^+_i u^+_j~.
\ee
It follows from (\ref{relations}) and the identity 
$[ D^{++}, D^+_{\hat \a} ]=0$,
that $\S$  and $\tilde{\L}^{++} $ are  analytic superfields, 
\be
D^+_{\hat \a} \S =0~, \qquad 
D^+_{\hat \a} \tilde{\L}^{++} =0~.
\ee
Representing
$\x = \x^{\hat a} \pa_{\hat a} 
-\x^{+\hat \a} D^-_{\hat \a} 
+ \x^{-\hat \a} D^+_{\hat \a}$,
one can now check that 
\be
[ \x -  \tilde{\L}^{++} D^{--} \, , \,  D^+_{\hat \a} ]
=  \tilde{\o}_{\hat \a}{}^{\hat \b}\, D_{\hat \b}^+
- (\S - 2\tilde{\s} ) \, D^+_{\hat \a}~.
\ee
This relation implies that the operator $ \x -  \tilde{\L}^{++} D^{--} $
maps every  analytic superfield into an analytic one.
It is worth pointing out that  the superconformal transformation of
the analytic subspace measure involves
\be
\pa_{\hat a} \x^{\hat a}  +D^-_{\hat \a} \x^{+\hat \a} 
-D^{--}\tilde{\L}^{++} =2\S~.
\ee

\subsection{Harmonic superconformal multiplets}
We present  here several superconformal multiplets
that are globally defined  over the harmonic superspace. 
Such a multiplet  is  described by a 
smooth Grassmann analytic superfields $\F^{(n)}_\k (z,u^+,u^-)$,
\be
D^+_{\hat \a}  \F^{(n)}_\k =0~,
\ee
which is endowed with  the  following superconformal transformation law
\be
\d \F^{(n)}_\k = - \Big(  \x -  \tilde{\L}^{++} D^{--} \Big) \, \F^{(n)}_\k 
-\k \,\S \, \F^{(n)}_\k ~.
\label{harmult1}
\ee
The parameter $\k$ is related to the conformal  weight of
$ \F^{(n)}_\k$. We will call $ \F^{(n)}_\k$ 
an analytic density  of weight $\k$.
When $ n$ is even, one can define 
real superfields, 
$\breve{\F}^{(n)}_\k=\F^{(n)}_\k$, 
with respect to the analyticity-preserving conjugation \cite{GIKOS,GIOS}
(also known as `smile-conjugation').

Let $V^{++}$ be a real analytic gauge potential describing 
a $U(1)$ vector multiplet. 
Its superconformal transformation is 
\be
\d V^{++} = - \Big(  \x -  \tilde{\L}^{++} D^{--} \Big) \, V^{++}~.
\label{v++tr}
\ee
Associated with the gauge potential is  the field strength 
\bea
W = \frac{\rm i}{8} \int {\rm d}u \, 
 ({\hat D}^-)^2 \, V^{++}~, \qquad 
 ({\hat D}^\pm)^2=D^{\pm \hat \a} D^\pm_{\hat \a}
 \label{W2}
\eea
which is  known to be invariant under the gauge transformation
$\d V^{++} = D^{++} \l $, where the gauge parameter
$\l$ is a real analytic superfield.
The  superconformal transformation of $W$,
 \bea
\d W = -\frac{\rm i}{8} \int {\rm d}u \, 
 ({\hat D}^-)^2  \Big(  \x + (D^{--} \tilde{\L}^{++})  \Big) \,
 V^{++}~,
\eea
can be shown to coincide with (\ref{vmfstransfo}). 

There are many ways to describe a hypermultiplet.
In particular, one can use 
an analytic superfield $q^+ (z,u)$ and its smile-conjugate
$\breve{q}^+ (z,u)$ \cite{GIKOS,GIOS}. They transform 
as follows:
\be
\d q^+ = - \Big(  \x -  \tilde{\L}^{++} D^{--} \Big) \, q^+ 
- \,\S \, q^+ ~, \qquad 
\d \breve{q}^+ = - \Big(  \x -  \tilde{\L}^{++} D^{--} \Big) \, \breve{q}^+ 
- \,\S \, \breve{q}^+ ~.
\label{q+-trlaw}
\ee

One has $\k =n$ in (\ref{harmult1}), if
the superfield is annihilated by  
$D^{++}$,   
\bea
&& D^+_{\hat \a}  H^{(n)} = D^{++}   H^{(n)} =0~ 
\quad \longrightarrow \quad
H^{(n)}(z,u) =  H^{i_1\dots i_n} (z) \,u^+_{i_1}  \dots u^+_{i_n} ~,
\non \\
&& \qquad \d H^{(n)} = - \Big(  \x -  \tilde{\L}^{++} D^{--} \Big) \, H^{(n)}
-n \,\S \, H^{(n)}~.
\label{O(n)-harm}
\eea
Here the harmonic-independent superfield 
$H^{i_1\dots i_n} $ transforms according to 
(\ref{lin1}) with $p=3n$. 

\subsection{Projective superconformal multiplets}
In the projective superspace approach, 
one  deals only with  superfields
${\bm \f}^{(n)} (z,u^+)$ obeying the constraints
\bea
&& D^+_{\hat \a}  {\bm \f}^{(n)} = D^{++}   {\bm \f}^{(n)} =0~, 
\qquad n\geq 0~.
\label{holom2}
\eea
Here the first constraint means that ${\bm \f}^{(n)} $ is Grassmann analytic, 
while the second constraint demands independence of $u^-$.
Unlike the harmonic superspace approach, however,
${\bm \f}^{(n)} (z,u^+)$ 
is not required to be well-defined over the two-sphere, that is, 
${\bm \f}^{(n)}$ may have singularities (say, poles)
at  some points of $S^2$. The presence of singularities 
turns out to be perfectly 
OK since the projective-superspace action involves 
a contour integral in $S^2$, see below.

We assume that ${\bm \f}^{(p)} (z,u)$  
is non-singular outside the north and south poles 
of $ S^2$. 
In the north chart,
we can represent 
\be
D^+_{\hat \a} = - u^{+\1}\,  \nabla_{\hat \a} (w)~, 
\qquad \nabla_{\hat \a} (w) = -D^i_{\hat \a} \, w_i~,
\qquad w_i = (-w, 1)~,
\label{nabla0}
\ee
Then, the equations
 (\ref{holom2})  
are equivalent to
\be
\f (z, w) = \sum_{n=-\infty}^{+\infty} \f_n (z) \,w^n~,
\qquad 
\nabla_{\hat \a} (w) \, \f(z,w)=0~,
\label{holom0}
\ee
with  the holomorphic superfield
$\f (z, w) \propto {\bm \f}^{(n)} (z,u^+)$.
These relations define a {\it projective multiplet}, following 
the four-dimensional terminology \cite{projective}.
Associated with $\f (z,w) $ is its smile-conjugate
\bea 
\breve{\f} (z, w) = \sum_{n=-\infty}^{+\infty} (-1)^n \,
{\bar  \f}_{-n} (z) \,w^n~, \qquad 
\nabla_{\hat \a} (w) \, \breve{\f}(z,w)=0~,
\label{holom3}
\eea
which is also a projective multiplet.
If $\breve{\f} (z, w) = {\f} (z, w) $, the projective superfield 
is called real. 

Below we present several superconformal multiplets
as defined in the north chart. The corresponding transformations 
laws involve the two analytic building blocks:
$$
\tilde{\L}^{++} (w)= \tilde{\L}^{ij} \,w^+_i w^+_j
=  \tilde{\L}^{\1 \1 }\, w^2 -2  \tilde{\L}^{\1 \2}\, w 
+ \tilde{\L}^{\2 \2} ~,\quad
\S (w) = \tilde{\L}^{\1 i} \,w_i +3 \tilde{\s}
=  - \tilde{\L}^{\1 \1} \,w + \tilde{\L}^{\1 \2} +3 \tilde{\s}~.
$$
Similar structures occur in the south chart, that is 
$$
\tilde{\L}^{++} (y)= \tilde{\L}^{ij} \,y^+_i y^+_j
=  \tilde{\L}^{\1 \1 } -2  \tilde{\L}^{\1 \2}\, y 
+ \tilde{\L}^{\2 \2} \,y^2~,\quad
\S (y) = \tilde{\L}^{\2 i} \,y_i +3 \tilde{\s}
=  - \tilde{\L}^{\1 \2}  + \tilde{\L}^{\2 \2}\, y +3 \tilde{\s}~.
$$
In the overlap of the two charts, we have
\bea
\tilde{\L}^{++} (y)&=& \frac{1}{w^2} \,\tilde{\L}^{++} (w)
\quad \longrightarrow \quad
 \tilde{\L}^{++} (y)\,\pa_y =- \tilde{\L}^{++} (w)\,\pa_w \non \\
\S(y) &=&
\S (w) +\frac{1}{w} \,\tilde{\L}^{++} (w)~.
\eea

To realise a massless  vector multiplet,
one uses the so-called tropical multiplet
described by  
\be 
V (z, w) = \sum_{n=-\infty}^{+\infty} 
V_n (z) \,w^n~, \qquad
\bar{V}_n  =  (-1)^n \,V_{-n}~.
\label{tropical}
\ee
Its superconformal transformation
\be
\d V= - \Big(  \x +  \tilde{\L}^{++} (w)\,\pa_w \Big) \, V~.
\label{tropicaltransf}
\ee
The field strength of the vector multiplet\footnote{A more
general form for the field strength (\ref{strength3}) is given
in Appendix B.} is
\be
W(z)  =- \frac{1}{ 16\pi {\rm i}} \oint {\rm d} w \,
(\hat{D}^-)^2  \, V(z,w) 
=\frac{1}{ 4 \pi {\rm i}} \oint \frac{{\rm d} w}{w} \,
\cP (w)   \, V(z,w) ~,
\label{strength3}
\ee
where
 \bea
\cP(w) =\frac{1}{ 4w} \, 
(\bar D_{\1})^2 + \pa_5 - \frac{w}{ 4} \, (D^{\1})^2~.
\label{Diamond}
\eea
The superconformal transformation of $W$
can be shown to coincide with (\ref{vmfstransfo}).
The field strength (\ref{strength3}) is invariant 
under the gauge transformation 
\be
\d V(z,w) = {\rm i}\Big( \breve{\l} (z,w)-\l (z,w) \Big)~,
\label{lambda4}
\ee
with $\l(z,w)$ an arbitrary arctic multiplet, 
see below.

To describe a massless off-shell hypermultiplet, one can use 
the so-called arctic  multiplet $\U (z, w)$:
\be
{\bm  q}^+ (z, u) =  u^{+\1}\, \U (z, w) \sim
\U (z, w)~, \quad \qquad
\U (z, w) = \sum_{n=0}^{\infty} \U_n (z) w^n~.
\label{qsingular}
\ee
The  smile-conjugation of $ {\bm  q}^+$ leads to the so-called
 the antarctic multiplet $\breve{\U} (z, w) $:
\be
\breve{{\bm q}}^+ (z, u) = u^{+\2} \,\breve{\U} (z, w) \sim
w\, \breve{\U} (z, w) \qquad \quad
\breve{\U} (z, w) = \sum_{n=0}^{\infty} (-1)^n {\bar \U}_n (z)
\frac{1}{w^n}\;.
\label{smileqsingular}
\ee
Their  superconformal transformations are
\bea
\d \U = - \Big(  \x &+&  \tilde{\L}^{++} (w)\,\pa_w \Big) \U 
- \S (w) \, \U ~, \non \\
 \d \breve{\U} =  
- \frac{1}{w}\Big(  \x &+&  \tilde{\L}^{++} (w) \,\pa_w \Big) (w\,\breve{\U} )
-\S (w) \,\breve{\U} ~.
\label{polarsuperconf}
\eea
In the south chart, these transformations take the form
\bea
\d \U = - \frac{1}{y} \Big(  \x &-&  \tilde{\L}^{++} (y)\,\pa_y \Big) (y\,\U )
- \S (y) \, \U ~, \non \\
 \d \breve{\U} =  
- \Big(  \x &-&  \tilde{\L}^{++} (y) \,\pa_y \Big) \breve{\U} 
-\S (y) \,\breve{\U} ~.
\eea
Both  $\U(z,w)$ and $\breve{\U}(z,w)$ constitute 
the so-called  polar multiplet.

Since the product of two arctic superfields is again arctic,
from (\ref{polarsuperconf}) we obtain more general transformation
laws
\bea
\d \U_\k = - \Big(  \x &+&  \tilde{\L}^{++} (w)\,\pa_w \Big) \U_\k 
- \k\,\S (w) \, \U_\k ~, \non \\
 \d \breve{\U}_\k =  
- \frac{1}{w^\k}\Big(  \x &+&  \tilde{\L}^{++} (w) \,\pa_w \Big) (w^\k\,\breve{\U}_\k )
-\k\,\S (w) \,\breve{\U}_\k ~,
\label{polarsuperconf-kappa}
\eea
for some parameter $\k$. 
The case $\k=1$ corresponds to free hypermultiplet dynamics, 
see below.

Since the product $U_\k =  \breve{\U}_\k \, \U_\k $ is a tropical multiplet, 
we obtain  more general transformation laws than the one 
defined by eq.  (\ref{tropicaltransf}):
\bea
 \d U_\k =  
- \frac{1}{w^\k}\Big(  \x &+&  \tilde{\L}^{++} (w) \,\pa_w \Big) (w^\k\,U_\k )
-2\k\,\S (w) \,U_\k ~.
\label{tropicaltransf-kappa}
\eea

${}$Finally, let  us consider the projective-superspace reformulation 
of the multiplets (\ref{O(n)-harm})  with an even superscript,
\bea
H^{(2n)} (z,u) &=& 
\big({\rm i}\, u^{+1} u^{+2}\big)^n H^{[2n]}(z,w) \sim
\big({\rm i}\, w\big)^n H^{[2n]}(z,w)~, \\
H^{[2n]}(z,w) &=&
\sum_{k=-n}^{n} H_k (z) w^n~,
\qquad  {\bar H}_k = (-1)^k H_{-k} ~. \non
\label{O(n)-proj}
\eea
The projective superfield $H^{[2n]}(z,w) $ is often called a real $O(2n)$ 
multiplet \cite{projective}.
Its superconformal transformation in the north chart is
\bea
 \d  H^{[2n]} &=&  
- \frac{1}{w^n}\Big(  \x +  \tilde{\L}^{++} (w) \,\pa_w \Big) (w^n\, H^{[2n]} )
-2n \,\S (w)\, H^{[2n]} ~.
\label{o2n}
\eea
In a similar way one can introduce complex $O(2n+1)$ 
multiplets. In what follows, we will use the same name
`$O(n)$ multiplet' for both harmonic multiplets (\ref{O(n)-harm}) 
and the projective ones just introduced.

Among the projective superconformal multiplets considered, it is only 
the $O(n)$ multiplets which can be lifted to well-defined representations
of the superconformal  group on a compactified 5D 
harmonic superspace. The other multiplets realise 
the superconformal algebra only.

\sect{5D superconformal theories}
\label{section:six}

With the tools developed,
we are prepared to constructing 5D superconformal theories.
Superfield  formulations for 
5D $\cN=1$ rigid supersymmetric theories 
were earlier elaborated in the harmonic  \cite{Z,KL}
and projective \cite{KL} superspace settings.\footnote{In
the case of 6D $\cN=(1,0)$ rigid supersymmetric theories, 
superfield formulations have been developed 
in the conventional \cite{6Dstand},
harmonic  \cite{6Dhar}
and projective \cite{6Dproj} superspace settings.}

\subsection{Models in harmonic superspace}

Let $\cL^{(+4)}$ be an analytic density of weight $+2$.
Its superconformal transformation is a total derivative,
\bea
\d \cL^{(+4)} &=& - \Big(  \x -  \tilde{\L}^{++} D^{--} \Big) \, \cL^{(+4)}
-2 \,\S \, \cL^{(+4)} \non \\ 
&=&-\pa_{\hat a} \Big( \x^{\hat a} \, \cL^{(+4)}\Big)
-D^-_{\hat \a} \Big( \x^{+ \hat \a} \, \cL^{(+4)}\Big)
+ D^{--} \Big( \tilde{\L}^{++} \, \cL^{(+4)}\Big)~.
\eea
Therefore, such a  superfield generates a superconformal invariant
of the form
\be
\int {\rm d} \z^{(-4)} \, \cL^{(+4)} ~,
\ee
where 
\be
\int {\rm d} \z^{(-4)}
:=
 \int{\rm d} u
\int {\rm d}^5 x  \, 
 (\hat{D}^-)^4  ~, \qquad 
(\hat{D}^\pm)^4 = -\frac{1}{ 32}  (\hat{D}^\pm)^2 
\,  (\hat{D}^\pm)^2~.
\ee
This is the harmonic superspace action \cite{GIOS} as applied 
to the five-dimensional case.

Let $V^{++}$ be the gauge potential of an Abelian 
vector multiplet.
Given a real $O(2)$ multiplet $\cL^{++}$,
\be
D^+_{\hat \a}  \cL^{++} = D^{++}  \cL^{++} =0~,\qquad
\d \cL^{++} = - \Big(  \x -  \tilde{\L}^{++} D^{--} \Big) \, \cL^{++}
-2 \,\S \, \cL^{++}~,
\label{tensor}
\ee 
we can generate the following superconformal invariant 
\be
\int {\rm d} \z^{(-4)} \, V^{++}\,\cL^{++} ~.
\ee
Because of the constraint $D^{++} \cL^{++} =0$, 
the integral is invariant under the vector multiplet 
gauge transformation $\d V^{++} =- D^{++} \l$, 
with $\l $ a real analytic gauge parameter.
 
The field strength of the vector multiplet, $W$, is a primary superfield 
with the transformation (\ref{vmfstransfo}). 
Using $W$, one can construct  
the following analytic superfield \cite{KL}
\be 
-{\rm i} \, G^{++} = 
D^{+ \hat \a} W \, D^+_{\hat \a} W 
+\hf \,W \,  ({\hat D}^+)^2 W ~, \qquad 
D^+_{\hat \a} G^{++}=D^{++}G^{++} =0 ~. 
\label{G++}
\ee
which transforms as a harmonic superfield weight 2,
\be
\d G^{++} = - \Big(  \x -  \tilde{\L}^{++} D^{--} \Big) \, G^{++}
-2 \,\S \, G^{++} ~.
\label{G++transf}
\ee
In other words, $G^{++}$ is a real $O(2)$ multiplet.
As a result,
the supersymmetric Chern-Simons action\footnote{A different form
for this action was given in \cite{Z}.} \cite{KL}  
\be
S_{\rm CS} [V^{++}]=  \frac{1}{12 }
\int {\rm d} \z^{(-4)} \, V^{++} \,
G^{++} ~
\label{CS2}
\ee
is superconformally invariant. 

Super Chern-Simons theory (\ref{CS2}) is quite remarkable 
as compared with  the superconformal models of a single vector 
multiplet in four and six dimensions. In the 4D $\cN=2$ case,
the analogue of $G^{++}$ in (\ref{CS2})  is known to be 
$D^{+\a} D^+_\a W= {\bar D}^+_{\dt \a} {\bar D}^{+\dt \a} {\bar W}$, 
with $W$ the chiral field strength, and therefore the model is free.
In the case 6D $\cN=(1,0)$, the analogue of $G^{++}$ in (\ref{CS2})  is
$(D^+)^4 D^-_{\hat \a} W^{-\hat \a}$, see \cite{ISZ}  for more details,
and therefore the models is not only free but also has higher derivatives.
It is only in five dimensions that the requirement of superconformal invariance  
leads to a nontrivial dynamical system.

The model  (\ref{CS2}) admits interesting generalisations.
In particular, given several Abelian vector multiplets $V^{++}_I$, 
where $I=1,\dots, n$,  the composite superfield (\ref{G++}) 
is generalised as follows: 
\bea
G^{++}  ~\to~ 
G^{++}_{IJ} =G^{++}_{(IJ)}
&=&{\rm i}\,
\Big\{ D^{+ \hat \a} W_{I} \, D^+_{\hat \a} W_{J} 
+\hf \,W_{(I} \,  ({\hat D}^+)^2 W_{J)} \Big\}~,  \non \\
D^+_{\hat \a} G^{++}_{IJ}&=&D^{++}G^{++}_{IJ} =0 ~.
\eea
The gauge-invariant and  superconformal action (\ref{CS2}) 
turns into 
\be
\tilde{S}_{\rm CS} =  \frac{1}{12 }
\int {\rm d} \z^{(-4)} \, V^{++}_I \, c_{I ,JK}\,
G^{++}_{JK} ~, 
\qquad
c_{I ,JK} =c_{I, KJ}~,
\label{CS3}
\ee
for some constant parameters $c_{I ,JK} $.
One can also generalise  the super Chern-Simons theory (\ref{CS2}) 
to the non-Abelian case.

In harmonic superspace, some superconformal transformation 
laws are effectively independent (if properly understood)
of the dimension of space-time. As a result, some 4D $\cN=2$ 
superconformal theories can be trivially extended to five dimensions.
In particular, the model for a massless 
$U(1)$ charged hypermultiplet \cite{GIKOS}
\be
\label{q-hyper}
S_{\rm hyper}= - \int  {\rm d} \zeta^{(-4)}\,
\breve{q}{}^+ \Big( D^{++} +{\rm i} \, e\, V^{++} \Big) \,q^+ 
\ee
can be seen to be  superconformal. This follows from 
eqs. (\ref{v++tr}) and 
(\ref{q+-trlaw}), in conjunction with the observation that
the transformation  laws of $q^+$ and 
$D^{++} q^+$ are identical.

The dynamical system $S_{\rm CS} + S_{\rm hyper}$ can be chosen 
to describe the supergravity compensator sector (vector multiplet plus
hypermultiplet) when describing 5D simple supergravity within 
the superconformal tensor calculus \cite{Ohashi,Bergshoeff}.
Then, the hypermultiplet charge $e$ is equivalent to the presence of 
a non-vanishing cosmological constant, similar to the 4D $\cN=2$ 
case  \cite{GIOS}.

Our next example is a naive 5D generalisation of the 4D $\cN=2$ 
 improved tensor multiplet  \cite{deWPV,LR,projective0}
which was described in the harmonic superspace approach in 
\cite{GIO1,GIOS}.
Let us  consider the action
\bea 
S_{\rm tensor} [H^{++}]
= \int {\rm d} \z^{(-4)} \,\cL^{(+4)} (H^{++}, u) ~,
\label{tensoraction1}
\eea
where
\bea
\cL^{(+4)} (H^{++}, u) = \m^3 \,
\Big( \frac{ \cH^{++} }{1 + \sqrt{ 1+ \cH^{++} \,c^{--}  }}
\Big)^2~, \qquad 
\cH^{++} = H^{++}
- c^{++} ~,
\eea 
with $\m$ a constant parameter of unit mass dimension, 
and  $c^{++}$ a (space-time) independent 
holomorphic vector field on $S^2$, 
\be
c^{\pm \pm }(u) = c^{ij} \,u^\pm_i u^\pm_j ~, \qquad 
c^{ij} c_{ij} =2~, \qquad c^{ij} = {\rm const}~.
\ee
Here $H^{++}(z,u)$ is a real $O(2)$ multiplet
possessing the superconformal transformation law 
(\ref{O(n)-harm}) with $n=2$.
The superconformal invariance of (\ref{tensoraction1}) 
can be proved in complete analogy to 
the detailed consideration given \cite{GIO1,GIOS}.

Now, let us couple the vector multiplet 
to the real  $O(2)$ multiplet
by putting forward  the action
\bea
S_{\rm vector-tensor}[V^{++},H^{++}]= 
S_{\rm CS} [V^{++}]
+ \kappa  \int {\rm d} \z^{(-4)} \, V^{++} \, H^{++}
+ S_{\rm tensor} [H^{++}]~,
\eea
with $\kappa$ a coupling constant.
This action is both gauge-invariant and superconformal. 
It is a five-dimensional generalisation of the 4D $\cN=2$ model 
for massive tensor multiplet introduced in \cite{Kuz-ten}.

The dynamical system $S_{\rm vector-tensor}$ can be chosen 
to describe the supergravity compensator sector (vector multiplet plus
tensor multiplet) when describing 5D simple supergravity within 
the superconformal tensor calculus \cite{Ohashi,Bergshoeff}.
Then, the coupling constant  $\kappa$ is equivalent to  
a cosmological constant, similar to the 4D $\cN=2$ 
case  \cite{BS}.

Finally, consider 
the vector multiplet model
\be
S_{\rm CS} [V^{++}]
+  S_{\rm tensor} [G^{++} / \m^3]~,
\ee
with $G^{++}$ the composite superfield (\ref{G++}). 
The second term here turns out to be  a unique superconformal 
extension of the $F^4$-term, where  $F$ is the field strength of the
component gauge field.  In this respect, it is instructive 
to recall its 4D $\cN=2$ analogue \cite{deWGR}
\be
\int {\rm d}^4 x \,{\rm d}^8 \q\, 
  \ln W \ln {\bar W}  ~.
\ee
The latter  can be shown \cite{BKT} to be a unique $\cN=2$ superconformal 
invariant  in the family of actions 
of the form $\int {\rm d}^4x \,{\rm d}^8 \q \,H(W, {\bar W})$
introduced for the first time in \cite{Hen}.
In five space-time dimensions, if one looks for a superconformal invariant 
of the form $\int {\rm d}^5x \,{\rm d}^8 \q \,H(W)$, the general solution 
is $H(W) \propto W$, as follows from (\ref{trmeasure1}) and (\ref{vmfstransfo}), 
and this choice corresponds to a total derivative.

\subsection{Models in projective superspace}

Let $\cL (z,w) $ be an analytic superfield transforming 
according  to 
(\ref{tropicaltransf-kappa})
 with $\k=1$. 
This transformation law can be rewritten as
\bea
w\,  \d \cL &=&  
-  \Big(  \x +  \tilde{\L}^{++} \,\pa_w \Big) (w \, \cL )
-2 w\, \S \, \cL \non \\
&=& -\pa_{\hat a} \Big( \x^{\hat a} \, w\, \cL \Big)
-D^-_{\hat \a} \Big( \x^{+ \hat \a} \, w\, \cL \Big)
-\pa_w  \Big( \tilde{\L}^{++} \, w\,\cL \Big)~.
\label{o2}
\eea
Such a superfield turns out to generate a
superconformal invariant of the form 
\bea
I =
\oint 
\frac{{\rm d} w}{2\pi {\rm i}}  \,
\int {\rm d}^5 x \,
(\hat{D}^-)^4  \, w\,\cL (z,w)~,
\label{projac1}
\eea
where 
 $\oint {\rm d} w $ is a (model-dependent) 
contour integral in ${\mathbb C}P^1$.
Indeed, it  follows from (\ref{o2}) that  this functional
does not change under the superconformal transformations.
Eq. (\ref{projac1})
generalises the projective superspace action \cite{projective0,Siegel}  
to the five-dimensional case.
A more general form for this action, which 
does not imply the projective gauge conditions (\ref{projectivegaugeN})
and is based on the construction in \cite{Siegel}, is given in Appendix B. 
  
It is possible to  bring the action (\ref{projac1}) to a somewhat  simpler form
if one exploits the fact that  $\cL$ is Grassmann analytic.
Using the considerations outlined in Appendix  C gives
\bea
\int {\rm d}^5 x \,
(\hat{D}^-)^4 \, \cL 
=\frac{1}{w^2} 
\int {\rm d}^5 x \,
D^4 \cL \Big|~, \qquad 
D^4 = \frac{1}{16}  (D^{\1})^2 ({\bar D}_{\1})^2 \Big|~.
\eea
Here $D^4$ is the Grassmann part of the integration measure of 4D $\cN=1$ 
superspace, 
$\int {\rm d}^4 \q = D^4$.
Then, functional (\ref{projac1}) turns into 
\bea
I= \oint  \frac{ {\rm d} w}{2\pi {\rm i} w} \,
\int {\rm d}^5 x \,
D^4 \cL 
= \oint  \frac{ {\rm d} w}{2\pi {\rm i}w} \int  {\rm d}^5 x \,{\rm d}^4 \q \, 
\cL ~.
\label{projac2}
\eea

Our first example is the tropical multiplet formulation 
for the super Chern-Simons theory \cite{KL}
\be
S_{\rm CS} = -
\frac{1}{12 } 
\oint 
\frac{{\rm d}w}{2\pi {\rm i}w} 
 \int {\rm d}^5 x \, 
{\rm d}^4 \q \,
V\,G ~,
\label{CS-proj}
\ee
with the contour around the origin. Here
$G(w) $ is the composite $O(2) $ multiplet 
(\ref{G++}) constructed out of the tropical gauge potential 
$V(w)$, 
\be
G^{++}= ({\rm i} \,u^{+\1}u^{+\2}) \, G(w)
 \sim {\rm i} \,w\,G(w)~,
\qquad G(w) = -\frac{1}{ w} \, \J+K+ w\,  \bar \J~,
\label{sYMRed}
\ee 
The explicit expressions for the superfields
$\J$ and $K$ 
can be found in \cite{KL}.
The above consideration 
and  the transformation laws 
(\ref{tropicaltransf}) and (\ref{G++transf}) imply that 
the action (\ref{CS-proj}) is superconformal.

Next, let us generalise to five dimensions 
the  charged $\U$-hypermultiplet model of \cite{projective}:
\be
S_{\rm hyper}= 
\oint \frac{{\rm d}w}{2\pi {\rm i}w} 
 \int {\rm d}^5 x \, 
{\rm d}^4 \q \,
\breve{\U}  \,{\rm e}^{ q \, V }\, \U ~,
\ee
with $q$  the hypermultiplet charge, and 
the integration contour around the origin. 
This action is superconformal, in accordance 
with the transformation laws (\ref{tropicaltransf})
and (\ref{polarsuperconf}).
It is also invariant under  gauge transformations
\be
\d \U = {\rm i} \, q \, \U ~, \qquad 
\d V = {\rm i} ( \breve{\l}-\l  )~,
\ee
with $\l$ an  arctic superfield.

Now, let us couple the vector multiplet to a real
$O(2)$  multiplet $H(w)$
\be
H^{++}= ({\rm i} \,u^{+\1}u^{+\2}) \, H(w)
 \sim {\rm i} \,w\,H(w)~,
\qquad H(w) = -\frac{1}{ w} \, \F+L + w\,  \bar \F~,
\label{O(2)-components}
\ee 
We introduce the vector-tensor system
\bea
S &=& -
\oint 
\frac{{\rm d}w}{2\pi {\rm i}w} 
 \int {\rm d}^5 x \, 
{\rm d}^4 \q \,
V \Big\{ \frac{1}{12 }\, G 
+\kappa \, H \Big\} 
+ \m^3 \oint 
\frac{{\rm d}w}{2\pi {\rm i}w} 
 \int {\rm d}^5 x \, 
{\rm d}^4 \q \, H \, \ln H ~, 
\label{vt-proj}
\eea
where the first term on the right involves a contour around the origin, 
while the second comes with a contour turning clockwise and anticlockwise
around the roots of of the quadratic equation 
$w\, H(w)=0$. The second term in (\ref{vt-proj}) is
a minimal 5D extension of the 4D $\cN=2$ improved tensor multiplet
\cite{projective0}.
It should be  pointed out that the component superfields 
in (\ref{O(2)-components}) obey the constraints \cite{KL}
\be 
{\bar D}^{\dt \a} \, \F =0~,
\qquad
-\frac{1}{ 4} {\bar D}^2 \, L 
= \pa_5\,  \F~. 
\ee 
It should be also remarked  that the real linear superfield $L$
can always be dualised into a chiral scalar and its conjugate \cite{KL}, 
which generates a special chiral superpotential. 

Given several $O(2) $ multiplets $H^I(w)$, where $I=1,\dots,n$,
superconformal dynamics is generated by the action
\be
S=\oint 
\frac{{\rm d}w}{2\pi {\rm i}w} 
 \int {\rm d}^5 x \, 
{\rm d}^4 \q \, \cF( H^I ) ~, \qquad I=1,\dots ,n~
\ee
where  $\cF (H) $ is a weakly homogeneous function 
of first degree in the variables $H$,
\be
 \oint \frac{{\rm d}w}{2\pi {\rm i}w}  \int {\rm d}^5 x \,  
{\rm d}^4 \q \,
\Big\{ H^I \, \frac{\pa \cF(H  ) }{\pa H^I} 
-\cF (H  ) \Big\} =0~.
\ee
This is completely analogous to the four-dimensional case
\cite{projective0,BS,dWRV} where the component structure
of such sigma-models has been studied in detail 
\cite{deWKV}.

A great many superconformal models can be obtained if one
considers $\U$-hypermultiplet actions of the form 
\bea
S = \oint \frac{{\rm d}w}{2\pi {\rm i}w} 
 \int {\rm d}^5 x \,  
{\rm d}^4 \q \,
K \big( \U^I  , \breve{\U}^{ \bar J}   \big)~,
 \qquad I,{\bar J} =1,\dots ,n~
\label{nact}
\eea
with the contour around the origin.
Let us first assume that the superconformal 
transformations of all $\U$'s and $\breve{\U}$'s 
have the form (\ref{polarsuperconf}).
Then,  in accordance with  general principles,
the action is superconformal if $K ( \U , \breve{\U} )  $ is 
a weakly homogeneous function of first degree in the variables $\U$,
\be
 \oint \frac{{\rm d}w}{2\pi {\rm i}w}  \int {\rm d}^5 x \,  
{\rm d}^4 \q \,
\Big\{ \U^I \, \frac{\pa K(\U, \breve{\U}  ) }{\pa \U^I} 
-K(\U, \breve{\U} ) \Big\} =0~.
\label{polar-homog}
\ee
This homogeneity condition is compatible
with the K\"ahler invariance
\be
K(\U, \breve{\U})  \quad \longrightarrow \quad K(\U, \breve{\U}) ~+~
\L(\U) \,+\, {\bar \L} (\breve{\U} ) 
\ee
which the model (\ref{nact}) possesses \cite{Kuzenko,GK,KL}.

Unlike  the $O(n)$ multiplets, the superconformal  transformations
of $\U$ and $\breve{\U}$ are not fixed uniquely by the constraints, 
as directly follows from (\ref{polarsuperconf-kappa}).
Therefore, one can consider superconformal sigma-models of the form
(\ref{nact}) in which the dynamical variables $\U$'s consist 
of several subsets with different values for the weight $\k$ in 
(\ref{polarsuperconf-kappa}), and then $K(\U, \breve{\U} )$
should obey weaker conditions than eq.  (\ref{polar-homog}).
Such a situation occurs, for instance,  if one starts with a gauged linear 
sigma-model and then integrates out the gauge multiplet, 
in the spirit of \cite{LR,dWRV}.
As an example, consider
\be
S= 
\oint \frac{{\rm d}w}{2\pi {\rm i}w} 
 \int {\rm d}^5 x \, 
{\rm d}^4 \q \,
\Big\{ 
\breve{\U}^\a \,\eta_{\a \b} \,   \U^\b  \,{\rm e}^{  V }
+ \breve{\U}^\m \,\eta_{\m \n} \,   \U^\n  \,{\rm e}^{ - V } \Big\} ~,
\ee
where $\eta_{\a \b} $ and $\eta_{\m \n}$ are constant 
diagonal metrics, 
$\a=1, \dots , m$ and $\m =1, \dots , n$.
Integrating out the tropical multiplet gives
the gauge-invariant action
\be
S= 2
\oint \frac{{\rm d}w}{2\pi {\rm i}w} 
 \int {\rm d}^5 x \, 
{\rm d}^4 \q \,
\sqrt{
\breve{\U}^\a \,\eta_{\a \b} \,   \U^\b  
\, \breve{\U}^\m \,\eta_{\m \n} \,   \U^\n  }~.
\ee
The gauge freedom can be completely fixed by setting, say, 
one of the superfields $\U^\n$ to be unity. 
Then, the action becomes
\be
S= 2
\oint \frac{{\rm d}w}{2\pi {\rm i}w} 
 \int {\rm d}^5 x \, 
{\rm d}^4 \q \,
\sqrt{
\breve{\U}^\a \,\eta_{\a \b} \,   \U^\b  
\,( \eta_{nn} + \breve{\U}^{\underline \m} \,
\eta_{\underline{\m} \underline{\n}} \,   
\U^{\underline \n})   }~,
\ee
where $\underline{\m}, \underline{\n}=1,\dots,n-1.$
This action is still superconformal, 
but  now $ \U^\b  $ and $\U^{\underline \n}$ transform 
according to (\ref{polarsuperconf-kappa})
with $\k=2$ and $\k=0$, respectively.

Sigma-models (\ref{nact}) have an interesting geometric interpretation 
if $K(\F, \bar \F )$ is the K\"ahler potential of a K\"ahler manifold
$\cM$  \cite{Kuzenko,GK,KL}.
Among the component superfields of 
$\U (z,w) = \sum_{n=0}^{\infty} \U_n (z) \,w^n$, 
the leading components 
$\F = \U_0 | $ and $\G = \U_1 |$ 
considered as 4D $\cN=1$ superfields, 
are constrained: 
\be 
{\bar D}^{\dt \a} \, \F =0~,
\qquad
-\frac{1}{ 4} {\bar D}^2 \, \G 
= \pa_5\,  \F~. 
\label{pm-constraints}
\ee 
The $\F$ and $\G$  can be regarded as 
a complex coordinate of the K\" ahler
manifold and a tangent vector at point $\F$ of the same manifold, 
and therefore they parametrize  the tangent bundle $T\cM$
of the K\"ahler manifold. 
The other components, $\U_2, \U_3, \dots$,
are complex unconstrained superfields. 
These superfields are auxiliary since they appear 
in the action without derivatives.
The auxiliary superfields $\U_2, \U_3, \dots$, and their
conjugates,  can
be eliminated  with the aid of the 
corresponding algebraic equations of motion
\be
 \oint {{\rm d} w} \,w^{n-1} \, \frac{\pa K(\U, \breve{\U} 
) }{\pa \U^I} = 0~,
\qquad n \geq 2 ~.               
\label{int}
\ee
Their elimination  can be  carried out
using the ansatz 
\bea
\U^I_n = \sum_{p=o}^{\infty} 
U^I{}_{J_1 \dots J_{n+p} \, \bar{L}_1 \dots  \bar{L}_p} (\F, {\bar \F})\,
\G^{J_1} \dots \G^{J_{n+p}} \,
{\bar 
\G}^{ {\bar L}_1 } \dots {\bar \G}^{ {\bar L}_p }~, 
\qquad n\geq 2~.
\eea
It can be shown that the coefficient functions 
$U$'s are uniquely determined by equations
(\ref{int}) in perturbation theory. 
Upon elimination of the auxiliary superfields,
the action 
(\ref{nact}) takes the form
\bea
S
[\F, \bar \F, \G, \bar \G]  
&=& \int {\rm d}^5 x \, 
{\rm d}^4 \q \,
 \Big\{\,
K \big( \F, \bar{\F} \big) - g_{I \bar{J}} \big( \F, \bar{\F} 
\big) \G^I {\bar \G}^{\bar{J}} 
\non\\
&&\qquad +
\sum_{p=2}^{\infty} \cR_{I_1 \cdots I_p {\bar J}_1 \cdots {\bar 
J}_p }  \big( \F, \bar{\F} \big) \G^{I_1} \dots \G^{I_p} {\bar 
\G}^{ {\bar J}_1 } \dots {\bar \G}^{ {\bar J}_p }~\Big\}~, 
\eea
where the tensors $\cR_{I_1 \cdots I_p {\bar J}_1 \cdots {\bar 
J}_p }$ are functions of the Riemann curvature $R_{I {\bar 
J} K {\bar L}} \big( \F, \bar{\F} \big) $ and its covariant 
derivatives.  Each term in the action contains equal powers
of $\G$ and $\bar \G$, since the original model (\ref{nact}) 
is invariant under rigid $U(1)$  transformations
\be
\U(w) ~~ \mapsto ~~ \U({\rm e}^{{\rm i} \a} w) 
\quad \Longleftrightarrow \quad 
\U_n(z) ~~ \mapsto ~~ {\rm e}^{{\rm i} n \a} \U_n(z) ~.
\label{rfiber}
\ee

The complex linear superfields $\G^I$ can be dualised 
into chiral superfields\footnote{This is accompanied 
by the appearance of a special chiral superpotential \cite{KL}.}  
$\J_I$ which can be interpreted as 
a one-form at the point $\F \in \cM$ \cite{GK,KL}. 
Upon elimination of $\G$ and $\bar \G$, 
the action turns into $S[\F, \bar \F, \J, \bar \J]$.
Its target space is  (an open neighborhood of the zero section) of 
the cotangent  bundle $T^*\cM$ of the K\"ahler manifold $\cM$. 
Since supersymmetry requires this target space to be hyper-K\"ahler, 
our consideration is in accord with recent mathematical results  
\cite{cotangent} about the existence of hyper-K\"ahler 
structures on cotangent  bundles of  K\"ahler manifolds.

\subsection{Models with intrinsic central charge}

We have so far considered only superconformal multiplets 
without central charge. As is known, there is no clash 
between superconformal symmetry 
and the presence of a central charge provided the latter is gauged.
Here we sketch a 5D superspace setting for supersymmetric theories 
with gauged central charge, which is a natural generalisation 
of the 4D $\cN=2$ formulation \cite{DIKST}.

To start with, one introduces an Abelian vector multiplet, which  
is destined to gauge the central charge $\D$, by defining 
gauge-covariant derivatives
\be
\cD_{\hat A} = ( \cD_{\hat a}, \cD_{\hat \a}^i ) 
= D_{\hat A} + \cV_{\hat A} (z)\, \D ~, \qquad 
[\D , \cD_{\hat A} ]=0~.
\ee
Here the gauge connection $  \cV_{\hat A} $
is inert under the central 
charge transformations, 
$[\D \,,\cV_{\hat A} ] =0$.
The gauge-covariant derivatives are required   
to obey the algebra
\bea
\{\cD^i_{\hat \a} \, ,  \cD^j_{\hat \b} \} &= &-2{\rm i} \,
\ve^{ij}\,
\Big( 
\cD_{\hat \a \hat \b} 
+ \ve_{\hat \a \hat \b} \, \cW \,\D \Big)~, 
\qquad \big[ \cD^i_{\hat \a} \, ,  \D \big] =0~,
\non \\
\big[ 
\cD^i_{\hat \g}\,, \cD_{\hat \a \hat \b} \big] &=& 
{\rm i}\,  
\ve_{\hat a \hat \b} \, \cD^i_{\hat \g}
 \cW\,\D
+2{\rm i}\,\Big( \ve_{\hat \g \hat \a} \,\cD^i_{\hat \b} 
-  \ve_{\hat \g \hat \b} \,\cD^i_{\hat \a} \Big)\cW \,\D
~, 
\label{SYM-algebra}
\eea
where  the real field strength $\cW(z)$ obeys the Bianchi identity
(\ref{Bianchi1}). The field strength should possess a non-vanishing 
expectation value, $\langle \cW \rangle \neq 0$,
corresponding to the case of rigid central charge.
By applying a harmonic-dependent gauge transformation, 
one can choose a frame in which
\be
\cD^+_{\hat \a} ~\to ~D^+_{\hat \a} ~, 
\quad D^{++} ~\to ~ D^{++} +\cV^{++}\,\D~,
\quad D^{--} ~\to ~ D^{--} +\cV^{--}\,\D~, 
\ee
with $\cV^{++} $ a real analytic prepotential, see \cite{DIKST}
for more details.
This frame is called the $\l$-frame, and the original representation 
is known as the $\t$-frame \cite{GIKOS}.

To generate a supersymmetric action, 
it is sufficient to construct a real superfield 
$\cL^{(ij)}(z)$ with the properties 
\be
\cD^{(i}_{\hat \a} \cL^{jk)} =0~,
\ee
which for $\cL^{++}(z,u) = \cL^{ij} (z) \, u^+_i u^+_j$ take the form
\be
\cD^+_{\hat \a} \cL^{++} = 0~, 
\qquad D^{++}\cL^{++} =0~.
\ee
In the $\l$-frame, the latter properties become
\be
D^+_{\hat \a} \tilde{\cL}^{++} = 0~, 
\qquad (D^{++} + \cV^{++} \,\D) \tilde{\cL}^{++} =0~.
\ee
Associated with $ \tilde{\cL}^{++}$ is 
the supersymmetric  action 
\be
 \int {\rm d} \z^{(-4)} \, \cV^{++}\,
\tilde{ \cL}^{++} 
\ee
which invariant under the  central charge gauge transformations
$\d \cV^{++} =- D^{++} \l $ and
$\d \tilde{\cL}^{++} = \l \,\D \, \tilde{\cL}^{++} $, 
with an arbitrary analytic parameter $\l$.

Let us give a few examples of off-shell supermultiplets 
with intrinsic central charge. The simplest is 
the 5D extension of the Fayet-Sohnius hypermultiplet.
It is described by an iso-spinor  superfield 
${\bm q}_i (z)$ 
and its conjugate ${\bar {\bm q}}^i (z)$
subject to the constraint 
\be
\cD^{(i}_{\hat \a} \, {\bm q}^{j)} =0~. 
\label{FSh}
\ee
This multiplet becomes on-shell if $\D = {\rm const}$.
With the notation ${\bm q}^+(z,u)  ={\bm q}^{j} (z) u^+_i$,
the hypermultiplet dynamics is dictated by the Lagrangian
\be
L^{++}_{\rm FS} = \hf \,
\breve{{\bm q}}^+ 
\stackrel{\longleftrightarrow}{ \D} 
{\bm q}^+ 
-{\rm i}\, m\, \breve{{\bm q}}^+ {\bm q}^+~,
\label{FS-Lagrangian}
\ee
with $m$ the hypermultiplet mass/charge.
This Lagrangian generates a superconformal theory.

Our second example is  an off-shell gauge  two-form multiplet 
called in \cite{Ohashi} the massless tensor multiplet.
It is Poincar\'e dual to the 5D vector multiplet.
Similarly to the 4D $\cN=2$ vector-tensor multipet \cite{DIKST},
it is described by a  constrained real superfield $L(z) $ coupled
to the central charge vector multiplet.  
By analogy with the four-dimensional  case \cite{DIKST}, 
admissible constraints must obey some 
nontrivial consistency conditions. In particular, 
the harmonic-independence of $L$ (in the $\t$-frame) 
implies 
\bea
0=(\hat{\cD}^+)^2 (\hat{\cD}^+)^2D^{--} L &=& D^{--} (\hat{\cD}^+)^2(\hat{\cD}^+)^2 L
-4 \,\cD^{-\hat \a} \cD^+_{\hat \a} (\hat{\cD}^+)^2L 
+8{\rm i}\, \cD^{\hat \a \hat \b} \cD^+_{\hat \a} \cD^+_{\hat \b}  \non \\
& - &8{\rm i}\, \D \,\Big\{ L \,(\hat{\cD}^+)^2 \cW 
+\cW \,(\hat{\cD}^+)^2 L +4\,\cD^{+\hat \a} \cW \, \cD^+_{\hat \a} L\Big\}~.
\label{consistency}
\eea
Let us assume that $L$ obeys the constraint
\be
\cD^+_{\hat \a} \cD^+_{\hat \b }  L
= \frac{1}{4} \ve_{\hat \a \hat \b} \,
({\hat \cD}^+)^2  L \quad 
\Rightarrow \quad 
\cD^+_{\hat \a} \cD_{\hat \b }^+  \cD_{\hat \g }^+ L
= 0
\label{Bianchi2}
\ee
which in the case $\D =0$ coincides with the Bianchi identity 
for an  Abelian vector multiplet. Then, eq. (\ref{consistency}) gives
\be
\D \, \Big\{ L \,(\hat{\cD}^+)^2 \cW 
+\cW \,(\hat{\cD}^+)^2 L +4\,\cD^{+\hat \a} \cW \, \cD^+_{\hat \a} L\Big\}
=0~.
\label{consistency2}
\ee

The consistency condition is satisfied if $L$ is constrained as
\be 
(\hat{\cD}^+)^2 L =- \frac{1}{\cW}\,\Big\{
 L \,(\hat{\cD}^+)^2 \cW 
+4\,\cD^{+\hat \a} \cW \, \cD^+_{\hat \a} L\Big\}~.
\label{two-form-constraint1}
\ee
The corresponding Lagrangian is 
\be
\cL^{++} = -\frac{\rm i}{4} \Big( \cD^{+ \hat \a}  L\,\cD^+_{\hat \a} L
+\hf \,L\,(\hat{\cD}^+)^2L\Big)~.
\label{two-form-lagrang}
\ee
The theory generated by this Lagrangian is superconformal.

Another solution to (\ref{consistency2}) describes a Chern-Simons
coupling of the two-form multiplet to an external Yang-Mills
supermultiplets:
\bea 
(\hat{\cD}^+)^2 L &=&- \frac{1}{\cW}\,\Big\{
 L \,(\hat{\cD}^+)^2 \cW 
+4\,\cD^{+\hat \a} \cW \, \cD^+_{\hat \a} L\Big\}
+ \frac{\r}{\cW}\,{\mathbb G}^{++}~,
\label{two-form-constraint2}
\eea
where 
\bea
-{\rm i} \, {\mathbb G}^{++} &=& {\rm tr}\,
\Big( \cD^{+ \hat \a} {\mathbb W} \, \cD^+_{\hat \a} {\mathbb W} 
+ \frac{1 }{ 4} \{ {\mathbb W} \,, 
({\hat \cD}^+)^2 {\mathbb W} \} \Big)~.
\eea
Here $\r$ is a  coupling constant, and $\mathbb W$
is the gauge-covariant field strength of the Yang-Mills 
supermultiplet,  see \cite{KL} for more details. 
As the corresponding supersymmetric Lagrangian 
one can  again choose $\cL^{++}$
given by eq. (\ref{two-form-lagrang}). 

A plain dimensional reduction $5{\rm D} \to 4{\rm D}$ 
can be shown to reduce the constraints 
(\ref{Bianchi2}) and (\ref{two-form-constraint2})
to those describing the so-called 
 linear vector-tensor multiplet\footnote{Ref. \cite{DIKST} 
contains an extensive  list of publications on 
the linear and nonlinear vector-tensor multiplets
and their couplings.}   
with Chern-Simons couplings
\cite{DIKST}.
\vskip.5cm

When this paper was ready for submission to the hep-th archive, 
there appeared an interesting work \cite{BX} 
in which 4D and 5D supersymmetric nonlinear 
sigma models with eight supercharges were formulated in 
$\cN=1$ superspace.

\noindent
{\bf Acknowledgements:}\\
It is a pleasure to thank Ian McArthur for reading the manuscript.
The author is grateful to the Max Planck Institute for Gravitational Physics
(Albert Einstein Institute) in  Golm
and  the Institute for Theoretical Physics at the
University of Heidelberg 
for hospitality during the course of the work.
This work  is supported  
by the Australian Research Council and by a UWA research grant.

\begin{appendix} 

\sect{Non-standard realisation for 
$\bm{ S^2}$  
} 

Let us consider a quantum-mechanical spin-$1/2$ Hilbert space, 
i.e. the complex space ${\mathbb C}^2$ endowed with 
the standard  positive definite scalar 
product $\langle ~|~\rangle$ defined by 
\bea
\langle u|v\rangle = u^\dagger \,v ={\bar u}^i \,v_i~, \qquad  
|u \rangle = (u_i) =\left(
\begin{array}{c}
u_1 \\
u_2  
\end{array}
\right)~,
\qquad \langle u | = ({\bar u}^i) ~,
\quad {\bar u}^i =\overline{u_i}~.
\eea
Two-sphere $S^2$ can be identified with the space 
of rays in ${\mathbb C}^2$. A ray is represented by a 
normalized state, 
\be
 |u^- \rangle = (u^-_i) ~,
\qquad 
\langle u^- | u^- \rangle=1~, 
\qquad \langle u^- | = (u^{+i}) ~,
\quad u^{+i} =\overline{u^-_i}~,
\ee
defined modulo the 
equivalence relation 
\be
u^-_i ~ \sim ~ {\rm e}^{ -{\rm i} \vf  } \, u^-_i~, 
\qquad | {\rm e}^{-{\rm i} \vf } |=1~.
\label{equivalence}
\ee

Associated with $|u^- \rangle $ is
another normalized state $|u^+ \rangle $, 
\bea
|u^+ \rangle = (u^+_i) ~,
\qquad 
u^+_i = \ve_{ij}\,u^{+j}~, \qquad
\langle u^+ | u^+ \rangle=1~, 
\eea
which is orthogonal to $|u^- \rangle $,
\be
\langle u^+ | u^- \rangle=0~.
\ee
The states $|u^- \rangle $  and $|u^+ \rangle $
generate the  unimodular unitary matrix
\bea
{\bm u}=\Big(  |u^- \rangle \, ,\,  |u^+ \rangle \Big)
=({u_i}^-\,,\,{u_i}^+) \in SU(2)~.
\eea
In terms of this matrix, 
the equivalence relation (\ref{equivalence}) becomes
\bea
{\bm u} ~\sim ~ {\bm u}\,
\left(
\begin{array}{cc}
 {\rm e}^{ -{\rm i} \vf  }  & 0\\
0&    {\rm e}^{ {\rm i} \vf  } 
\end{array}
\right)~.
\eea
This gives
the well-known realisation 
$S^2 = SU(2) /U(1)$.

The above unitary realisation for $S^2$ is ideal if
one is interested in the action of $SU(2)$, or its subgroups,
on the two-sphere. 
But it is hardly convenient if one considers, 
for instance,  the action of $SL(2,{\mathbb C})$  on $S^2$.
There exists, however,  a universal realisation.
Instead of dealing with the orthonormal basis 
$(  |u^- \rangle ,  |u^+ \rangle )$ introduced, 
one can work with an arbitrary basis for ${\mathbb C }^2$:
\bea
{\bm v}=\Big(  |v^- \rangle \, ,\,  |v^+ \rangle \Big)
=({v_i}^-\,,\,{v_i}^+) \in GL(2,{\mathbb C})~,\qquad 
\det {\bm v}=v^{+i}\,v^-_i~. 
\eea
The two-sphere is then obtained by factorisation with respect to 
the equivalence relation
\bea
{\bm v} ~\sim ~ {\bm v}\,R~,
\qquad 
R= \left(
\begin{array}{cc}
 a  & 0\\
b &      c 
\end{array}
\right) \in GL(2,{\mathbb C})~.
\label{equivalence2}
\eea
Given an arbitrary  matrix ${\bm v} \in  GL(2,{\mathbb C})$, 
there always exists a lower triangular matrix $R$ such that 
${\bm v} R \in SU(2)$, 
and then we are back to the unitary realisation.  
One can also consider an intermediate realisation for $S^2$  
given in terms of  unimodular matrices of the form 
\bea
{\bm w}=\Big(  |w^- \rangle \, ,\,  |w^+ \rangle \Big)
=({w_i}^-\,,\,{w_i}^+) \in SL(2,{\mathbb C}) \quad 
\longleftrightarrow \quad w^{+i} w^-_i =1~,
\eea
and the matrix $R$ in (\ref{equivalence2}) should be 
restricted to be unimodular.
The harmonics $w^\pm$ are complex in the sense that 
$w^-_i$ and $w^{+i}$ are not related by complex conjugation.

Let us consider a left group transformation acting on $S^2$
\bea
{\bm w} ~\to ~g\, {\bm w}= ({v_i}^-\,,\,{v_i}^+) \equiv {\bm v}~.
\eea
If $g$ is a ``small''  transformation, i.e. if it belongs 
to a sufficiently small neighbourhood of the identity, 
then there exists a matrix $R$ of the type (\ref{equivalence2}) 
such that 
\be
g\, {\bm w} \,R = ({w_i}^-\,,\,{\hat{w}_i}^+) \in SL(2,{\mathbb C}) ~.
\ee
Since
$$
w^{+i} w^-_i =1~,\qquad \hat{w}^{+i} {w}^-_i =1~,
$$
for the transformed harmonic we thus obtain 
\be
\hat{w}^+_i = w^+_i  + \r^{++}(w) \,w^-_i ~.
\ee
All information about the group transformation $g$ 
is now encoded in $ \r^{++} $.

\sect{Projective superspace action}
${}$Following  \cite{Siegel}, consider
\bea
I= \frac{1}{2\pi {\rm i}}  \oint
\frac{ u^+_i\,{\rm d} u^{+i}}{(u^+ u^-)^4}  \,
\int {\rm d}^5 x \,
(\hat{D}^-)^4  \,\cL^{++} (z,u^+)~,
\label{projac3}
\eea
where the Lagrangian enjoys the properties
\be
D^+_{\hat \a} \cL^{++} (z,u^+)=0~,
\qquad 
\cL^{++} (z,c\,u^+) = c^2\, \cL^{++} (z,u^+)~, 
\qquad c \in{\mathbb C}^*~.
\ee
The functional (\ref{projac3}) 
is invariant under arbitrary projective transformations
(\ref{equivalence22}).
Choosing the projective gauge (\ref{projectivegaugeN})
gives the supersymmetric action (\ref{projac1}). 

It is worth pointing out that 
the vector multiplet field strength (\ref{strength3})
can be rewritten in the projective-invariant  form 
\be
W(z)  =- \frac{1}{ 16\pi {\rm i}} \oint 
\frac{ u^+_i\,{\rm d} u^{+i}}{(u^+ u^-)^2}  \,
(\hat{D}^-)^2  \, V(z,u^+)~, 
\label{strengt4}
\ee
where the gauge potential enjoys the properties
\be
D^+_{\hat \a} V (z,u^+)=0~,
\qquad 
V (z,c\,u^+) = V (z,u^+)~,\qquad c \in{\mathbb C}^*~.
\ee

\sect{From 5D projective supermultiplets to 
4D 
$\bm{ \cN=1, \,2}$ 
superfields}

The conventional 5D simple superspace ${\mathbb R}^{5|8}$ 
is parametrized  
by  coordinates  $ z^{\hat A} = (x^{\hat a},  \q^{\hat \a}_i )$, 
with $i = \1 , \2$. 
Any hypersurface $x^5 ={\rm const}$ in ${\mathbb R}^{5|8}$ 
can be identified  with the 4D, $\cN=2$  superspace 
${\mathbb R}^{4|8}$ parametrized by 
$
z^{A} = (x^a,  \q^\a_i , {\bar \q}_{\dt \a}^i)$,  
where $(\q^\a_i )^* = {\bar \q}^{\dt \a i}$.
The Grassmann coordinates of ${\mathbb R}^{5|8}$ and 
${\mathbb R}^{4|8}$
are related to each other as follows:
\bea
\q^{\hat \a}_i = ( \q^\a_i , - {\bar \q}_{\dt \a i})~, 
\qquad
\q_{\hat \a}^i =   
\left(
\begin{array}{c}
\q_\a^i \\
{\bar \q}^{\dt \a i}    
\end{array}
\right)~.
\eea
Interpreting $x^5$ as a central charge variable, 
one can view ${\mathbb R}^{5|8}$ as a 4D, $\cN=2$ 
central charge superspace.
One can relate the 5D spinor covariant derivatives
(see \cite{KL} for more details)
\bea
D^i_{\hat \a} 
= \left(
\begin{array}{c}
D_\a^i \\
{\bar D}^{\dt \a i}    
\end{array}
\right)
=D^i_{\hat \a} = \frac{\pa}{\pa \q^{\hat \a}_i} 
- {\rm i} \, (\G^{\hat b} ){}_{\hat \a \hat \b} \, \q^{\hat \b i}
\, \pa_{\hat b}
~, 
\qquad 
D^{\hat \a}_i  = 
(D^\a_i \,, \, -{\bar D}_{\dt \a i}) 
\label{con}
\eea 
to the 4D, $\cN=2$ covariant derivatives 
$D_A = (\pa_a , D^i_\a , {\bar D}^{\dt \a}_i )$
 where 
\bea
 D^i_\a &=& \frac{\pa}{\pa \q^{\a}_i}
+ {\rm i} \,(\s^b )_{\a \bd} \, {\bar \q}^{\dt \b i}\, \pa_b
+
\q^i_\a \, \pa_5 ~,
\quad 
{\bar D}_{\dt \a i} = 
- \frac{\pa}{\pa {\bar \q}^{\dt \a i}} 
- {\rm i} \, 
 \q^\b _i (\s^b )_{\b \dt \a} \,\pa_b
-{\bar \q}_{\dt \a i} \,
\pa_5 ~. 
\label{4D-N2covder1}
\eea
These operators obey the anti-commutation relations
\bea
\{D^i_{\a} \, , \, D^j_{ \b} \} = 2 \,
\ve^{ij}\, \ve_{\a \b} \,
\pa_5 ~,
\quad && \quad 
\{{\bar D}_{\dt \a i} \, , \, {\bar D}_{\dt  \b j} \} = 2 \,
\ve_{ij}\, \ve_{\dt \a \dt \b} \,
\pa_5 ~, 
\non \\
\{D^i_{\a} \, , \, \bar D_{ \dt \b j} \} &=& -2{\rm i} \, \d^i_j\,
(\s^c )_{\a \dt \b} \,\pa_c ~,
\label{4D-N2covder2}
\eea
which correspond to  the 4D, $\cN=2$ supersymmetry algebra with 
the central charge $\pa_5$.

Consider a 5D projective superfield (\ref{holom0}).
Representing the differential operators 
$\nabla_{\hat \a} (w)$, eq.  (\ref{nabla0}), as
\bea
\nabla_{\hat \a} (w) 
= \left(
\begin{array}{c}
\nabla_\a (w) \\
{\bar \nabla}^{\dt \a}  (w)  
\end{array}
\right)~, 
\quad \nabla_\a (w) \equiv  w D^{\1}_\a - D^{\2}_\a ~,
\quad
{\bar \nabla}^{\dt \a} (w) \equiv {\bar D}^{\dt \a}_{ \1} + 
w {\bar D}^{\dt \a}_{ \2}~,
\label{nabla}
\eea
the constraints (\ref{holom3})
can be rewritten in the component form
\be
D^{\2}_\a \f_n = D^{\1}_\a \f_{n-1} ~,\qquad
{\bar D}^{\dt \a}_{\2} \f_n = - {\bar D}^{\dt \a}_{ \1} 
\f_{n+1}~.
\label{pc2}
\ee
The relations (\ref{pc2}) imply  that the dependence
of the component superfields 
$\f_n$ on $\q^\a_{\2}$ and ${\bar \q}^{\2}_{\dt \a}$ 
is uniquely determined in terms 
of their dependence on $\q^\a_{\1}$
and ${\bar \q}^{\1}_{\dt \a}$.  In other words, 
the projective superfields depend effectively 
on half the Grassmann variables which can be choosen
to be the spinor  coordinates of 4D $\cN=1$ superspace
\be
\q^\a = \q^\a_{\1} ~, \qquad {\bar \q}_{\dt \a}=
{\bar \q}_{\dt \a}^{\1}~.
\label{theta1}
\ee 
Then,  one deals with  reduced   superfields
$\f | $, $ D^{\2}_\a \f|$, $ {\bar D}_{\2}^{\dt \a} \f|, \dots$
(of which not all are usually independent)    
and 4D $\cN=1$ spinor covariant derivatives $D_\a$ 
and ${\bar D}^{\dt \a}$ defined in the obvious way:
\be 
\f| = \f (x, \q^\a_i, {\bar \q}^i_{\dt \a})
\Big|_{ \q_{\2} = {\bar \q}^{\2}=0 }~,
\qquad D_\a = D^{\1}_\a \Big|_{\q_{\2} ={\bar \q}^{\2}=0} ~, 
\qquad
{\bar D}^{\dt \a} = {\bar D}_{\1}^{\dt \a} 
\Big|_{\q_{\2} ={\bar \q}^{\2}=0}~.
\label{N=1proj}
\ee

\end{appendix}

\small{

}


\begin{thebibliography}{66}

\bi{Nahm}  W.~Nahm,
 ``Supersymmetries and their representations,''
  Nucl.\ Phys.\ B {\bf 135} (1978) 149.


\bibitem{Kac}
  V.~G.~Kac,
  ``Lie superalgebras,''
  Adv.\ Math.\  {\bf 26} (1977) 8.

\bibitem{Ohashi}
  T.~Kugo and K.~Ohashi,
  ``Off-shell d = 5 supergravity coupled to matter-Yang-Mills system,''
  Prog.\ Theor.\ Phys.\  {\bf 105} (2001) 323
{[hep-ph/0010288]};
  T.~Fujita and K.~Ohashi,
  ``Superconformal tensor calculus in five dimensions,''
  Prog.\ Theor.\ Phys.\  {\bf 106} (2001) 221
 {[hep-th/0104130]};
 T.~Fujita, T.~Kugo and K.~Ohashi,
  ``Off-shell formulation of supergravity on orbifold,''
  Prog.\ Theor.\ Phys.\  {\bf 106}  (2001) 671
{[hep-th/0106051]};
   T.~Kugo and K.~Ohashi,
  ``Superconformal tensor calculus on orbifold in 5D,''
  Prog.\ Theor.\ Phys.\  {\bf 108}  (2002) 203
{[arXiv:hep-th/0203276]}.

\bibitem{Bergshoeff}
  E.~Bergshoeff, S.~Cucu, T.~De Wit, J.~Gheerardyn, R.~Halbersma, 
  S.~Vandoren and A.~Van Proeyen,
  ``Superconformal N = 2, D = 5 matter with and without actions,''
  JHEP {\bf 0210}  (2002) 045
 {[hep-th/0205230]};
E.~Bergshoeff, S.~Cucu, T.~de Wit, J.~Gheerardyn, S.~Vandoren and A.~Van Proeyen,
  ``N = 2 supergravity in five dimensions revisited,''
  Class.\ Quant.\ Grav.\  {\bf 21}  (2004) 3015
{[arXiv:hep-th/0403045]}. 

\bibitem{GIKOS}
A.~Galperin, E.~Ivanov, S.~Kalitsyn, V.~Ogievetsky 
and E.~Sokatchev,
``Unconstrained N = 2 matter, Yang-Mills 
and supergravity theories 
in harmonic superspace,''
{Class.\ Quant.\ Grav.\  {\bf 1}  (1984)} 469.

\bibitem{GIOS}
A.~S.~Galperin, E.~A.~Ivanov, V.~I.~Ogievetsky and E.~S.~Sokatchev,
{\it Harmonic Superspace}, Cambridge University Press, 
Cambridge, 2001.

\bi{projective0}
 A.~Karlhede, U.~Lindstr\"om and M.~Ro\v{c}ek,
  ``Self-interacting tensor multiplets in N = 2 superspace,''
  Phys.\ Lett.\ B {\bf 147}  (1984) 297.

\bibitem{Siegel}
  W.~Siegel,
 ``Chiral actions for N=2 supersymmetric tensor multiplets,''
  Phys.\ Lett.\ B {\bf 153} (1985) 51.

\bibitem{projective}
 U.~Lindstr\"om and M.~Ro\v{c}ek,
  ``New hyperkahler metrics and new supermultiplets,''
  Commun.\ Math.\ Phys.\  {\bf 115}  (1988) 21;
``N = 2 super Yang-Mills theory in projective superspace,''
  Commun.\ Math.\ Phys.\  {\bf 128}  (1990) 191;
 F.~Gonzalez-Rey, M.~Ro\v{c}ek, S.~Wiles, 
U.~Lindstr\"om  and R.~von Unge,
  ``Feynman rules in N = 2 projective superspace. 
I: Massless  hypermultiplets,''
  Nucl.\ Phys.\ B {\bf 516}  (1998) 426
{[hep-th/9710250]}.

\bibitem{BS}
  N.~Berkovits and W.~Siegel,
  ``Superspace effective actions for 4D compactifications of heterotic 
and type  II superstrings,''
  Nucl.\ Phys.\ B {\bf 462} (1996) 213 [hep-th/9510106];
  N.~Berkovits,
  ``Conformal compensators and manifest type IIB S-duality,''
  Phys.\ Lett.\ B {\bf 423} (1998) 265
  [hep-th/9801009].

\bi{Rosly}
 A.~A.~Rosly,
``Super Yang-Mills  constraints 
as integrability conditions,'' in {\it Group Theoretical 
Methods in Physics},'' M. A. Markov  (Ed.), 
Nauka, Moscow, 1983, p. 263.

\bi{RS} A.~A.~Rosly and A.~S.~Schwarz,
  ``Supersymmetry in a space with auxiliary dimensions,''
 Commun.\ Math.\ Phys.\  {\bf 105}, 645 (1986).

\bibitem{Witten}
  E.~Witten,
  ``An interpretation of classical Yang-Mills theory,''
  Phys.\ Lett.\ B {\bf 77} (1978) 394.

\bibitem{Kuzenko}
S.~M.~Kuzenko,
``Projective superspace as a double-punctured harmonic superspace,''
Int.\ J.\ Mod.\ Phys.\ A {\bf 14} (1999) 1737
{[hep-th/9806147]}.

\bi{KL}
 S.~M.~Kuzenko and W.~D.~Linch,
``On five-dimensional superspaces,''
hep-th/0507176.


\bibitem{dWRV}
  B.~de Wit, M.~Ro\v{c}ek and S.~Vandoren,
  ``Hypermultiplets, hyperkaehler cones and quaternion-Kaehler geometry,''
  JHEP {\bf 0102} (2001) 039 [hep-th/0101161].

\bi{GIOS-conf} 
A.~S.~Galperin, E.~A.~Ivanov, V.~I.~Ogievetsky and E.~S.~Sokatchev,
``Conformal invariance in harmonic superspace,'' 
in {\it Quantum Field Theory and Quantum Statistics},
I. Batalin, C. J. Isham and G. Vilkovisky (Eds.),
Vol. 2, Adam Hilger, Bristol, 1987.  

\bibitem{ISZ}
  E.~A.~Ivanov, A.~V.~Smilga and B.~M.~Zupnik,
  ``Renormalizable supersymmetric gauge theory in six dimensions,''
  Nucl.\ Phys.\ B {\bf 726} (2005) 131 [hep-th/0505082].

\bibitem{Rosly2}
  A.~A.~Rosly,
 ``Gauge fields in superspace and twistors,''
  Class.\ Quant.\ Grav.\  {\bf 2} (1985) 693.

\bibitem{LN}
  J.~Lukierski and A.~Nowicki,
  ``General superspaces from supertwistors,''
  Phys.\ Lett.\ B {\bf 211} (1988) 276.

\bibitem{HH}
  P.~S.~Howe and G.~G.~Hartwell,
  ``A superspace survey,''
  Class.\ Quant.\ Grav.\  {\bf 12} (1995) 1823.

\bi{Manin} Yu. I. Manin, {\it Gauge Field Theory and Complex Geometry},
Springer, Berlin, 1988.

\bibitem{twistors}
R.~Penrose, 
``Twistor algebra,''   J.\ Math.\ Phys.\  {\bf 8} (1967) 345;
R.~Penrose and M.~A.~H.~MacCallum,
 ``Twistor theory: An approach to the quantization 
of fields and space-time,''
 Phys.\ Rept.\  {\bf 6} (1972) 241.

\bi{WW} R. S. Ward and R. O. Wells, 
{\it Twistor Geometry and Field Theory}, 
Cambridge University Press, Cambridge, 1991.

\bibitem{GS}
  G.~W.~Gibbons and A.~R.~Steif,
  ``Sphalerons and conformally compactified Minkowski space-time,''
  Phys.\ Lett.\ B {\bf 346} (1995) 255
  [hep-ph/9412210].

\bi{S} I. E. Segal,  {\it Mathematical Cosmology and Extragalactic
Astronomy}, Academic Press, New York, 1976.


\bibitem{AGMOO}
  O.~Aharony, S.~S.~Gubser, J.~M.~Maldacena, H.~Ooguri and Y.~Oz,
  ``Large N field theories, string theory and gravity,''
  Phys.\ Rept.\  {\bf 323} (2000) 183
  [hep-th/9905111].

\bibitem{Sohnius}
  M.~F.~Sohnius,
 ``The conformal group in superspace,''
 in {\it Quantum Theory and the Structures of Time and Space}, Vol. 2,  
L. Castell , M. Drieschner and  C. F. von Weizs\"acker (Eds.), 
Carl Hanser Verlag, M\"unchen, 1977,  p. 241.
 
\bibitem{Lang}
  W.~Lang,
  ``Construction of the minimal superspace translation tensor and the
  derivation of the supercurrent,''
  Nucl.\ Phys.\ B {\bf 179} (1981) 106.

\bibitem{BPT}
  L.~Bonora, P.~Pasti and M.~Tonin,
``Cohomologies and anomalies in supersymmetric theories,''
  Nucl.\ Phys.\ B {\bf 252} (1985) 458. 

\bibitem{Shizuya}
  K.~i.~Shizuya,
  ``Supercurrents and superconformal symmetry,''
  Phys.\ Rev.\ D {\bf 35} (1987) 1848.

\bibitem{BK} I.~L.~Buchbinder and S.~M.~Kuzenko,
{\it Ideas and Methods of Supersymmetry and
Supergravity or a Walk Through Superspace},
IOP, Bristol, 1998.

\bibitem{West}
  P.~C.~West,
  ``Introduction to rigid supersymmetric theories,''
 hep-th/9805055. 
 
\bibitem{Osborn}
  H.~Osborn,   ``N = 1 superconformal symmetry in four-dimensional 
quantum field theory,''
Annals Phys.\  {\bf 272} (1999) 243 [hep-th/9808041].

\bibitem{Park}
  J.~H.~Park,
 ``Superconformal symmetry in six-dimensions and its reduction to four,''
  Nucl.\ Phys.\ B {\bf 539} (1999) 599
  [hep-th/9807186];
 ``Superconformal symmetry and correlation functions,''
  Nucl.\ Phys.\ B {\bf 559} (1999) 455
  [hep-th/9903230].

\bi{KT}
S.~M.~Kuzenko and S.~Theisen,
  ``Correlation functions of conserved currents in N = 2 superconformal
  theory,''
  Class.\ Quant.\ Grav.\  {\bf 17} (2000) 665
  [hep-th/9907107].


\bi{U} A.~Uhlmann, ``The closure of Minkowski space,''
Acta Phys. Pol. {\bf 24} (1963) 295.

\bi{Tod} I. T. Todorov, M. C. Mintchev and V. P. Petkova,
{\it Conformal Invariance in Quantum Field Theory},
Pisa, Scuola Normale Superiore, 1978;
I. T. Todorov,  {\it Conformal Description of Spinning Particles},
Springer, Berlin, 1986.

\bi{W} H. Weyl, {\it Space--Time--Matter}, Dover Publications, New York, 1952.

\bi{D} P. A. M. Dirac, ``Wave equations in conformal space,''
Ann. Math. {\bf 37} (1936) 429.


\bi{Cartan} \'E. Cartan, ``Sur les domaines born\'es 
homog\`enes de l'espace de $n$ variables
complexes,'' Abh. Math. Sem. Univ. Hamburg
{\bf 11} (1935) 116.

\bi{Hua} L. K. Hua, {\it Harmonic Analysis of Functions of 
Several Complex Variables in the Classical Domains}, 
American Mathematical Society, Providence, 1963.

\bi{PS} S. M. Paneitz and I. E. Segal, 
``Analysis in space-time bundles. I. General considerations 
and the scalar bundle,''  J. Funct. Anal. {\bf 47} (1982) 78.

\bibitem{WB} J.~Wess and J.~Bagger,
{\it Supersymmetry and Supergravity},
Princeton Univ. Press, 1992.

\bibitem{Ferber}
  A.~Ferber, ``Supertwistors and conformal supersymmetry,''
  Nucl.\ Phys.\ B {\bf 132} (1978) 55.

\bibitem{DeWitt}
  B.~S.~DeWitt,
  {\it Supermanifolds},
Cambridge University Press, Cambridge, 1992.


\bibitem{Z}
  B.~Zupnik,
  ``Harmonic superpotentials and symmetries in gauge theories 
with eight  supercharges,''
Nucl.\ Phys.\ B {\bf 554} (1999) 365
{[hep-th/9902038]}.


\bibitem{6Dstand}
P.~S.~Howe, G.~Sierra and P.~K.~Townsend,
  ``Supersymmetry in six dimensions,''
  Nucl.\ Phys.\ B {\bf 221} (1983) 331.

\bi{6Dhar} 
P.~S.~Howe, K.~S.~Stelle and P.~C.~West,
  ``N=1 D = 6 harmonic superspace,''
  Class.\ Quant.\ Grav.\  {\bf 2} (1985) 815;
B.~M.~Zupnik,
  Sov.\ J.\ Nucl.\ Phys.\  {\bf 44} (1986) 512.


\bibitem{6Dproj}
J.~Grundberg and U.~Lindstrom,
  ``Actions for linear multiplets in six dimensions,''
  Class.\ Quant.\ Grav.\  {\bf 2} (1985) L33;
  S.~J.~Gates, S.~Penati and G.~Tartaglino-Mazzucchelli,
  ``6D supersymmetry, projective superspace and 4D, N = 1 superfields,''
 hep-th/0508187.


\bibitem{deWPV}
B.~de Wit, R.~Philippe and A.~Van Proeyen,
``The improved tensor multiplet in N = 2 supergravity,''
Nucl.\ Phys.\ B {\bf 219} (1983) 143.

\bibitem{LR}
U.~Lindstrom and M.~Ro\v{c}ek,
``Scalar tensor duality and N = 1, 2 nonlinear sigma models,''
Nucl.\ Phys.\ B {\bf 222} (1983) 285.


\bibitem{GIO1}
A.~Galperin, E.~Ivanov and V.~Ogievetsky,
``Superspace actions and duality transformations 
for N = 2 tensor multiplets,''
Sov.\ J.\ Nucl.\ Phys.\  {\bf 45} (1987) 157;
``Duality transformations and most general matter 
self-coupling in N = 2 supersymmetry,''
Nucl.\ Phys.\ B {\bf 282} (1987) 74.

\bibitem{Kuz-ten}
S.~M.~Kuzenko,
 ``On massive tensor multiplets,''
 JHEP {\bf 0501} (2005) 041
 [hep-th/0412190].

\bibitem{deWGR}
B.~de Wit, M.~T.~Grisaru and M.~Ro\v{c}ek,
``Nonholomorphic corrections to the one-loop 
N=2 super Yang-Mills action,''
Phys.\ Lett.\ B {\bf 374} (1996) 297
[hep-th/9601115].

\bibitem{BKT}
  I.~L.~Buchbinder, S.~M.~Kuzenko and A.~A.~Tseytlin,
  ``On low-energy effective actions in N = 2,4 superconformal theories 
in  four dimensions,''
Phys.\ Rev.\ D {\bf 62} (2000) 045001 [hep-th/9911221].

\bibitem{Hen}
M.~Henningson,
``Extended superspace, higher derivatives 
and SL(2,Z) duality,''
Nucl.\ Phys.\ B {\bf 458} (1996) 445
[hep-th/9507135].

\bibitem{GK}
  S.~J.~Gates and S.~M.~Kuzenko,
  ``The CNM-hypermultiplet nexus,''
  Nucl.\ Phys.\ B {\bf 543} (1999) 122
  [hep-th/9810137];
 ``4D N = 2 supersymmetric off-shell sigma models on the cotangent  bundles of
  K\"ahler manifolds,''
  Fortsch.\ Phys.\  {\bf 48} (2000) 115
  [hep-th/9903013].

\bibitem{DIKST}
  N.~Dragon, E.~Ivanov, S.~Kuzenko, E.~Sokatchev and U.~Theis,
  ``N = 2 rigid supersymmetry with gauged central charge,''
  Nucl.\ Phys.\ B {\bf 538} (1999) 411
  [hep-th/9805152].
 
 \bibitem{deWKV}
  B.~de Wit, B.~Kleijn and S.~Vandoren,
  ``Rigid N = 2 superconformal hypermultiplets,''
 in {\it Supersymmetries and Quantum Symmetries},
 J. Wess and E. A. Ivanov (Eds.), Springer, Berlin, 1999, p. 37,
hep-th/9808160;
  ``Superconformal hypermultiplets,''
  Nucl.\ Phys.\ B {\bf 568} (2000) 475
  [hep-th/9909228].
 
\bibitem{cotangent} D.~ Kaledin, ``Hyperk\"ahler 
structures on total spaces of holomorphic cotangent bundles,''
in D. Kaledin and M. Verbitsky, {\it Hyperk\"ahler Manifolds},
International Press, Cambridge MA, 1999
[alg-geom/9710026]; 
B. Feix, ``Hyperk\"ahler metrics on  cotangent bundles,''
Cambridge PhD thesis, 1999;
J. reine angew. Math. {\bf 532}, 33 (2001).

\bi{BX} J. Bagger and C. Xiong, 
``N=2 nonlinear sigma models in N=1 superspace: Four and 
five dimensions,'' hep-th/0601165.


\end{thebibliography}
\end{document}